\documentclass[%
 reprint,
superscriptaddress,
 amsmath,amssymb,
aps,
]{revtex4-2}

\usepackage{graphicx}
\usepackage{amssymb}
\usepackage{amsmath}
\usepackage{dcolumn}
\usepackage{bm}
\usepackage{braket}
\usepackage{xcolor}
\usepackage{todonotes}
\usepackage{hyperref}
\hypersetup{
    colorlinks=true,
    linkcolor=blue,
    filecolor=magenta,      
    urlcolor=red,
    citecolor=blue,
}

\usepackage{xcolor}

\usepackage{soul}

\begin{document}
\title{Distinct orbital contributions to electronic and magnetic structures in La$_{4}$Ni$_{3}$O$_{10}$}

\author{Shilong Zhang}\thanks{These authors contributed equally to this work.}
\affiliation{International Center for Quantum Materials, School of Physics, Peking University, Beijing 100871, China}

\author{Hengyuang Zhang}\thanks{These authors contributed equally to this work.}
\affiliation{Guangdong Provincial Key Laboratory of Magnetoelectric Physics and Devices, School of Physics, Sun Yat-Sen University, Guangzhou, 510275, China }

\author{Zehao Dong}
\affiliation{State Key Laboratory of Low Dimensional Quantum Physics, Department of Physics, Tsinghua University, Beijing 100084, China}

\author{Jie Li}
\affiliation{National Laboratory of Solid State Microstructures and Department of Physics, Nanjing University, Nanjing 210093, China}

\author{Qian Xiao}
\affiliation{International Center for Quantum Materials, School of Physics, Peking University, Beijing 100871, China}
\affiliation{School of Physics and Information Engineering, Guangdong University of Education, Guangzhou 510303, China}

\author{Mengwu Huo}
\affiliation{Guangdong Provincial Key Laboratory of Magnetoelectric Physics and Devices, School of Physics, Sun Yat-Sen University, Guangzhou, 510275, China }

\author{Hsiao-Yu Huang}
\affiliation{National Synchrotron Radiation Research Center, Hsinchu 30076, Taiwan}

\author{Di-Jing Huang}
\affiliation{National Synchrotron Radiation Research Center, Hsinchu 30076, Taiwan}

\author{Yayu Wang}
\affiliation{State Key Laboratory of Low Dimensional Quantum Physics, Department of Physics, Tsinghua University, Beijing 100084, China}

\author{Yi Lu}
\affiliation{National Laboratory of Solid State Microstructures and Department of Physics, Nanjing University, Nanjing 210093, China}

\author{Zhen Chen}
\affiliation{Beijing National Laboratory for Condensed Matter Physics, Institute of Physics, Chinese Academy of Sciences, Beijing 100190, China}

\author{Meng Wang}
\email{wangmeng5@mail.sysu.edu.cn}
\affiliation{Guangdong Provincial Key Laboratory of Magnetoelectric Physics and Devices, School of Physics, Sun Yat-Sen University, Guangzhou, 510275, China }

\author{Yingying Peng}
\email{yingying.peng@pku.edu.cn}
\affiliation{International Center for Quantum Materials, School of Physics, Peking University, Beijing 100871, China}
\affiliation{Collaborative Innovation Center of Quantum Matter, Beijing 100871, China}

\date{\today} 

\begin{abstract}
High-$T_c$ superconductivity has recently been discovered in Ruddlesden --Popper phase nickelates under pressure, where the low-energy electronic structure is dominated by Ni $d_{x^2 - y^2}$
and $d_{z^2}$ orbitals.
However, the respective roles of these orbitals in superconductivity remain unclear. Here, by combining X-ray absorption, electron energy loss spectroscopy, and density functional theory calculations on La$_{4}$Ni$_{3}$O$_{10}$ single crystals, we identify ligand holes in the $p_{x,y}$ orbitals of planar oxygen and the $p_z$ orbitals of apical oxygen, which hybridize with the Ni $d_{x^2-y^2}$ and $d_{z^2}$ orbitals, respectively. These ligand holes enable orbital-selective O $K$-edge resonant inelastic X-ray scattering (RIXS) study, which reveals that $d_{x^2-y^2}$ states dominate the low-energy charge excitations and are more itinerant.
We also observe a $\sim$0.1 eV bimagnon through RIXS and Raman spectroscopy, which leads to an interlayer superexchange interaction $J_z$ of $\sim$50 meV. Our results reveal distinct contributions of Ni $d_{x^2-y^2}$ and $d_{z^2}$ orbitals to the electronic and magnetic structure and provide direct experimental insights to understand the RP-phase nickelate superconductors.
\end{abstract}

\maketitle

Recently, superconductivity has been discovered in bilayer and trilayer 
Ruddlesden--Popper (RP) nickelate single crystals under pressure, with maximum 
transition temperatures $T_c$ of about 80~K and 30~K, respectively~\cite{LNO327_SC_WM_Nat,LPNO327_SC_CJG_nat,LNO327_SC_CJG_PRX,LNO4310_SC_ZJ_Nat,PNO4310_SC_WM_CPL,LNO4310_SC_WHH_CPL,LNO4310_SC_doping_Japan}. Remarkably, ambient-pressure superconductivity ($T_c \approx 40$~K) has been realized 
in ultrathin LPNO films via strain engineering~\cite{LNO327film_SC_Nat,LPNOfilm_SC_XQK_Nat}. These RP-phase nickelates possess 3d$^7$ to 3d$^8$ electron occupancy and additional interlayer apical oxygen atoms, which distinguish them from the infinite-layer nickelate superconductors RNiO$_2$~\cite{RNO112_SC_Nat,RNO112_SC_linearR_nat,RNO112_SC_PRX2015,RNO112_SC_doping_PRL}. 
To understand the superconducting mechanism, extensive research on RP-phase nickelates has investigated their macroscopic and microscopic electronic and magnetic structures. Angle-resolved photoelectron spectroscopy (ARPES) reveals a multiband structure, with density functional theory (DFT) calculations attributing these bands to Ni 3$d_{x^2 - y^2}$ and 3$d_{z^2}$ orbitals~\cite{LNO327_band_ZXJ_NC,LNO4310_ARPES_band_NC,LNO327_Theory_band_PRB2023}. Experiments indicate that RP-phase nickelates exhibit a spin density wave (SDW) state at ambient pressure, which vanishes under pressure~\cite{LNO327_RIXS_NC,LNO4310_SDW_NC,LNO_NMR_SDW,LNO327_SDW_FastR_presure_NC,LNO327_SDW_R}. Resonant inelastic X-ray scattering (RIXS) at the Ni $L$ edge has observed dispersive magnon up to 70 meV and 50 meV in bilayer and trilayer RP-phase nickelates, respectively~\cite{LNO327_RIXS_NC,LNO4310_RIXS_PRB}. 
Various theoretical models have been proposed to explain the superconducting pairing mechanism~\cite{LNO327_Theory_pair_PRL2023,LNO327_Theory_pair_SDW_NC2024,LNO327_Theory_press_PRB2023,LNO327_Theory_ele_struct_PRL2023,LNO327_theory_YDX_npj2024,LNO4310_theory_SC_PRB2024,LNO4310_theory_SC_PRB2025,LNO4310_theory_interlay_PRB2024,LNO327_theory_dxy_pair_PRL2024,LNO327_theory_dxy_pair_2_PRL2024,LNO327_Theory_dxy_pair_PRB2024,LNO4310_Theory_dz_pair_PRB2024,LNO327_theory_spinsiglet_PRB2023}. These theories suggest that interlayer hopping is essential to the pairing mechanism, but there remains a key open question  as to whether superconducting pairing originates from the $d_{z^2}$ or the $d_{x^2 - y^2}$ orbitals. Therefore, it is essential to experimentally identify the electronic states of the $d_{z^2}$ and $d_{x^2-y^2}$ orbitals to understand RP-phase nicklate superconductors.

However, distinguishing the contributions of the Ni 3$d_{z^2}$ and 3$d_{x^2 - y^2}$ orbitals remains challenging, as both exhibit a substantial density of states near the Fermi energy. In Ni L$_3$-edge RIXS measurements, electrons can be excited into the Ni 3$d_{z^2}$ and 3$d_{x^2 - y^2}$ orbitals simultaneously, obscuring the orbital origin of the observed magnetic excitations. 
Previous electron energy loss spectroscopy (EELS) studies on La$_3$Ni$_2$O$_7$ have revealed ligand holes in both planar and apical oxygens~\cite{EELS_327_Nat,LPNO_EELS_Dong_NM}. These states can hybridize separately to the Ni 3$d_{x^2 - y^2}$ and 3$d_{z^2}$ orbitals, offering an indirect route to probe these orbitals individually. Moreover, employing different polarizations in the RIXS study, it is possible to excite electrons into different O 2p orbitals~\cite{XAS_pol_1996}, providing orbital-sensitive information in the energy loss spectra~\cite{RIXS_REXS_tech_1}. 

In this work, we combine O $K$-edge X-ray absorption spectroscopy (XAS), RIXS, EELS, and Raman spectroscopy to investigate a tri-layer RP-phase nickelate La$_{4}$Ni$_{3}$O$_{10}$ single crystal~\cite{LNO4310_struct_pre_WM}. Our results reveal that the ligand holes reside in the planar O $p_{x,y}$ orbitals and the apical O $p_z$ orbitals via hybridization with the Ni 3$d_{x^2 - y^2}$ and 3$d_{z^2}$ orbitals, respectively. 
Further investigation of excited states reveals that the $d_{x^2 - y^2}$ orbitals dominate low-energy charge excitations and are more itinerant than the $d_{z^2}$ orbitals.
We also observed a $\sim$0.1 eV bimagnon from the intralayer spin exchange, which leads to an interlayer superexchange interaction J$_z$ of $\sim$50 meV. Raman measurements at varying temperatures show that the bimagnon weakens significantly with increasing temperature but does not disappear above $T_{\mathrm{SDW}}$, indicating robust short-range 
magnetic fluctuations. These results reveal the distinct orbital contributions to the electronic and magnetic structure and advance our understanding of the RP-phase nicklate superconductors.

\begin{figure}[htbp]
    \centering
    \includegraphics[width = \columnwidth]{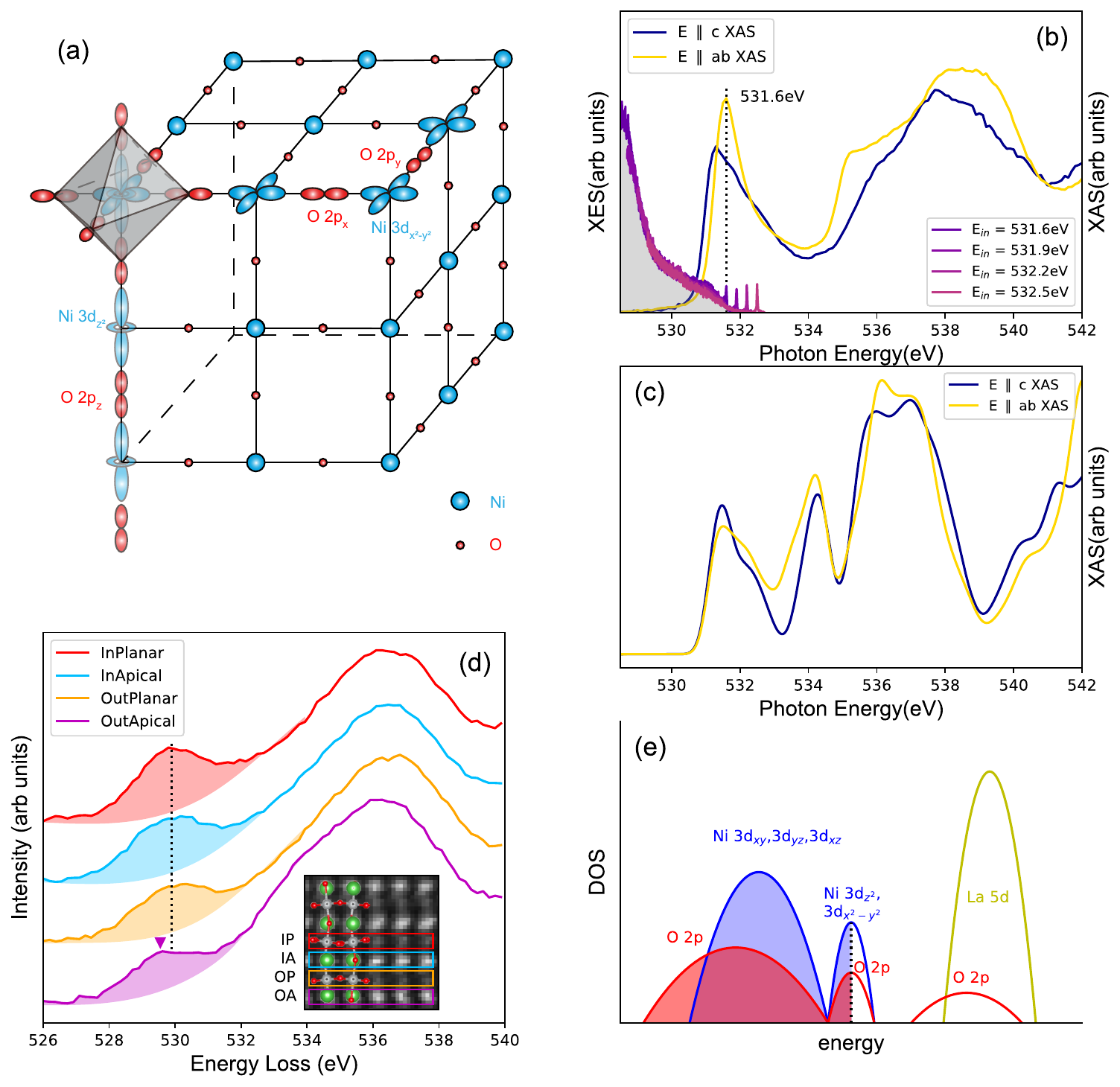}
    \caption{
        \label{fig1} 
        \textbf{(a)} Schematic diagram of the atomic orbitals near the Fermi energy of La$_{4}$Ni$_{3}$O$_{10}$.
        \textbf{(b)} The XAS spectra at O $K$-edge along in-plane ab direction and c direction and XES spectra obtained from the RIXS spectra. The dashed line marks the Fermi level obtained by XES. These indicate that both O $p_{x,y}$ and $p_z$ orbitals have ligand holes at the Fermi energy.
        \textbf{(c)} Calculated XAS spectra along in-plane and out-of-plane directions, which indicate that the holes are distributed in the orbitals shown in (a).
        \textbf{(d)} Occupancy of oxygen holes at different positions obtained by EELS. The small energy discrepancy in the pre-peak compared with the XAS feature is due to different instrumental calibrations.
        \textbf{(e)} Schematic plot of DOS in La$_{4}$Ni$_{3}$O$_{10}$, showing that the oxygen 2$p$ orbital is partially filled.
        }
\end{figure}

To probe the oxygen states in La$_4$Ni$_3$O$_{10}$ near the Fermi level, we performed O K-edge XAS and RIXS measurements. The trilayer structure is illustrated in Fig.~\ref{fig1}(a), using the P\textit{I}/mmm lattice~\cite{LNO4310_struct_pre_WM} with $a \approx b \approx 3.9~\text{\AA}$ and $c \approx 27.9~\text{\AA}$; reciprocal lattice units (r.l.u.) are defined accordingly, and the scattering plane lies in the $ac$ plane. By measuring the X-ray absorption spectra at different incident angles and polarizations, we decomposed the absorption spectra of La$_{4}$Ni$_{3}$O$_{10}$ to in-plane $ab$- and out-of-plane $c$-directions, as shown in Fig.~\ref{fig1}(b). The XAS spectra show strong prepeak in both directions, indicating that holes are present in the O $p_x$, $p_y$, and $p_z$ orbitals. 
We measured O $K$-edge RIXS spectra at different incident x-ray energies and obtained fluorescence excitation spectra, which are analogous to X-ray emission spectroscopy (XES) spectra and provide information on occupied states. 
The cutoff energy of these fluorescence excitations identifies the Fermi energy and aligns with the XAS pre-peak, suggesting that the oxygen holes are located at the Fermi level~\cite{LNO112_band_XASXES_NM2020}.

We also performed XAS calculations for in-plane and out-of-plane directions (Fig.~\ref{fig1}(c)). It generally reproduces the experimental line shapes, with the discrepancy reflecting the well-known tendency of DFT-based O $K$-edge simulations to produce a compressed energy scale relative to the experiment~\cite{DFT_XAS_mismath_1,DFT_XAS_mismath_2}. The calculations confirm that the prepeak in the oxygen $K$-edge arises from the orbitals shown in Fig.~\ref{fig1}(a).
Our O $K$-edge EELS measurements on La$_{4}$Ni$_{3}$O$_{10}$ reveal the spatial distribution of oxygen holes, as shown in Fig.~\ref{fig1}(d). Prepeaks are observed in both planar and inner apical oxygens, consistent with previous results on La$_3$Ni$_2$O$_7$~\cite{EELS_327_Nat}. A weaker and lower-energy prepeak appears in the outer apical oxygen, differing from  La$_3$Ni$_2$O$_7$. 
Based on these results, we constructed a schematic diagram of the energy state distribution in Fig.~\ref{fig1}(e). Near the Fermi energy, in addition to the Ni $d_{x^2-y^2}$ and $d_{z^2}$ orbitals, there are also ligand hole states in the oxygen 2p orbitals. These ligand hole states are located on the planar oxygen $p_{x,y}$ orbitals and the apical oxygen $p_z$ orbitals, which hybridize with the intralayer Ni 3$d_{x^2 - y^2}$ orbitals and the interlayer Ni 3$d_{z^2}$ orbitals, respectively. These ligand holes enable the independent probing of the 3$d_{x^2 - y^2}$ and 3$d_{z^2}$ orbitals by separately exciting distinct oxygen atomic orbitals.

\begin{figure}[htbp]
    \centering
    \includegraphics[width = \columnwidth]{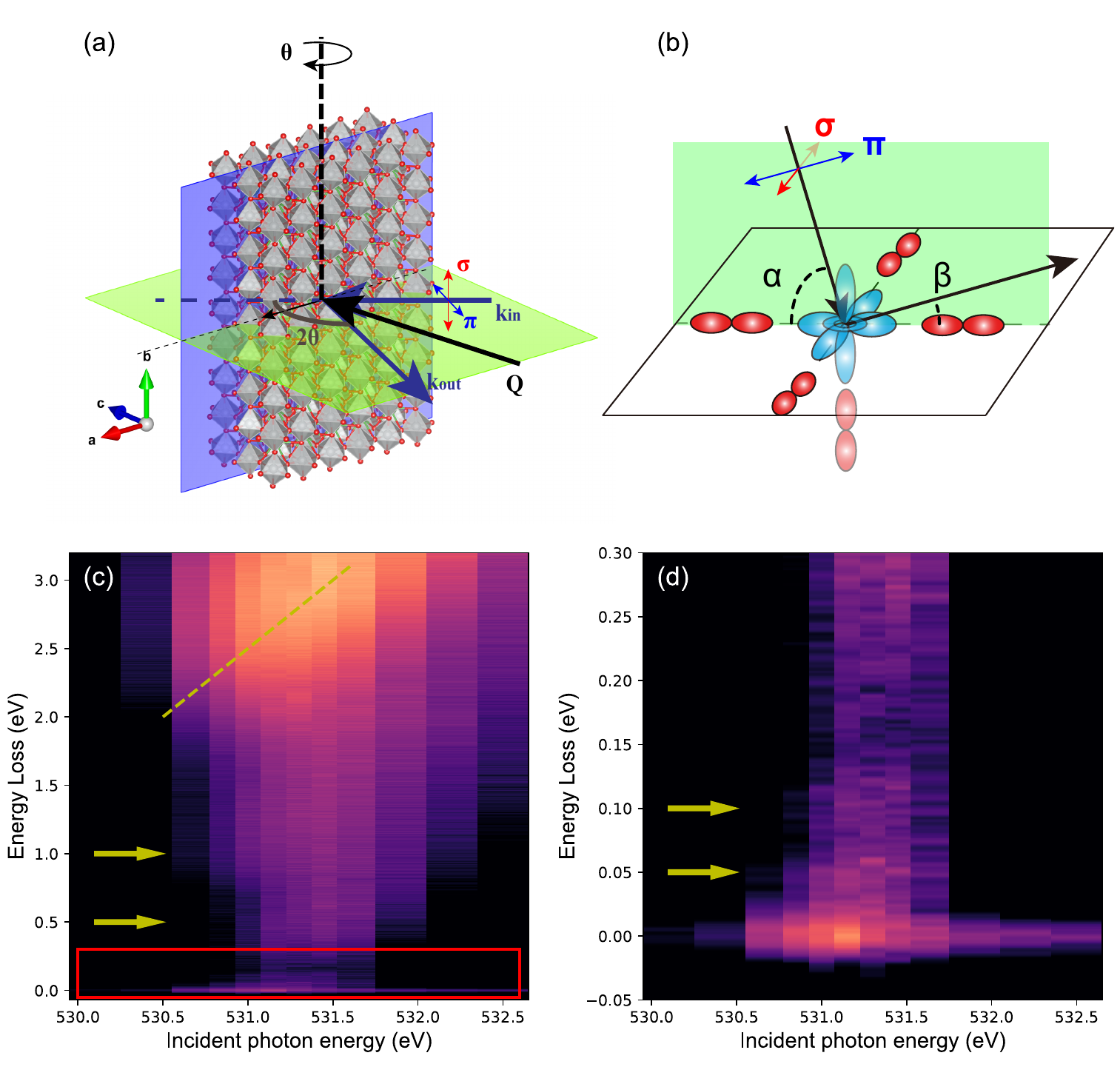}
    \caption{
        \label{fig2} 
        \textbf{(a)} Schematic plot of scattering geometry with lattice orientation. We define the ab-plane as the planar Ni-O plane and a-direction along the Ni-O-Ni direction, while c denotes the direction perpendicular to the planar Ni-O planes.
        \textbf{(b)} Schematic plot of scattering plane with different polarizations.
        \textbf{(c)} Incident energy detuning map of RIXS, where two Raman features are found at about 0.5 eV and 1 eV.
        \textbf{(d)} Zoom in of the red boxed region in (c) showing low-energy Raman excitations at around 50 meV and 100 meV.
        }
\end{figure}

 We then performed RIXS measurement with different incident x-ray polarizations to identify excitations from in-plane and out-of-plane O p orbitals. The experimental geometry is shown in Fig.~\ref{fig2}(a) and polarizations are shown in Fig.~\ref{fig2}(b). The $\sigma$-polarization direction (perpendicular to the scattering plane) always aligned within the $ab$ plane of the sample, while the $\pi$-polarization direction (parallel to the scattering plane) has an out-of-plane projection depending on the incident angle.
By scanning the incident x-ray energy at $\pi$ polarization, we identify several features with constant energy loss: two high-energy excitations at $\sim$0.5 eV and $\sim$1.0 eV (Fig.~\ref{fig2}(c)), and two low-energy modes at $\sim$50 meV and $\sim$100 meV (Fig.~\ref{fig2}(d)).

\begin{figure}[htbp]
    \centering
    \includegraphics[width = \columnwidth]{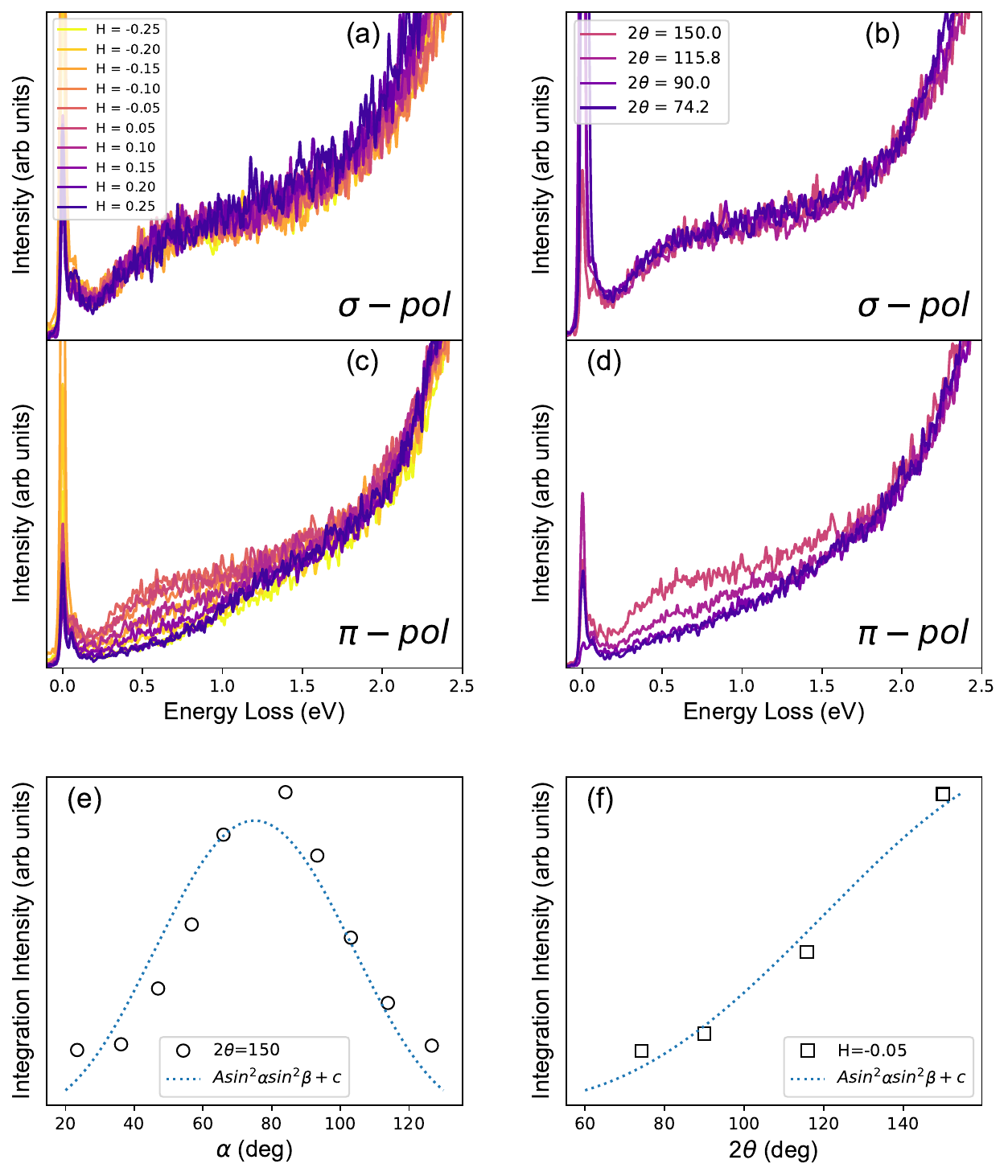}
    \caption{
        \label{fig3} 
        \textbf{(a,c)} H-dependent RIXS spectra at $\sigma$ and $\pi$ incident polarization, where the scattering angle $2\theta$ is fixed to 150 deg. 
        \textbf{(b, d)} $2\theta$-dependent RIXS spectra at $\sigma$ and $\pi$ incident polarization, where the H is fixed to 0.05 r.l.u.
        \textbf{(e, f)} The integrated intensity near 0.5 eV, extracted from the RIXS spectra in \textbf{(c,d)}, which follow $A sin^{2}\alpha sin^{2}\beta+c$ behavior.
        }
\end{figure}

To determine their orbital character, we then performed $H$-dependent and $2\theta$-dependent RIXS measurements. For $H$-dependent studies, the scattering angle $2\theta$ was fixed at $150^\circ$, while varying the in-plane momentum transfer $H$ along a-axis direction. For $2\theta$-dependent measurements, we fixed $H = -0.05 \ \text{r.l.u.}$ while scanning $2\theta$. We first focus on the high-energy spectra and found a pronounce 0.5 eV excitation with distinct polarization dependence. Its intensity remains constant under $\sigma$-polarization (Fig.~\ref{fig3}(a,b)) but shows significant changes under $\pi$-polarization (Fig.~\ref{fig3}(c,d)). The integration of the $\pi$ polarized spectral weight reveals an angular dependence well described by $A \sin^{2}\alpha \sin^{2}\beta + c$ (Fig.~\ref{fig3}(e,f)), where $\alpha$ and $\beta$ represent the incoming and outgoing angles (see Fig.~\ref{fig2}(b)). We notice that for charge excitations originating from in-plane orbitals, the $\pi-\pi'$ channel will exhibit a polarization factor of $\sin^{2}\alpha \sin^{2}\beta$, while the $\sigma-\sigma'$ channel remains constant. This behavior matches our results, and based on the electronic structure revealed in Fig.~\ref{fig1}(a), we then assign this excitation to a charge excitation with O $p_{x,y}$ and Ni $d_{x^2 - y^2}$ orbital components.
We observed a less pronounced feature above 1eV, which can be observed only under grazing incident angle with $\pi$-polarization conditions (Fig.~\ref{fig1}(c)), suggesting its origin from out-of-plane orbitals. The different behaviors between the in-plane components and the out-of-plane components suggest that these two orbitals do not hybridize together.

\begin{figure*}[htbp]
    \centering
    \includegraphics[width = \textwidth]{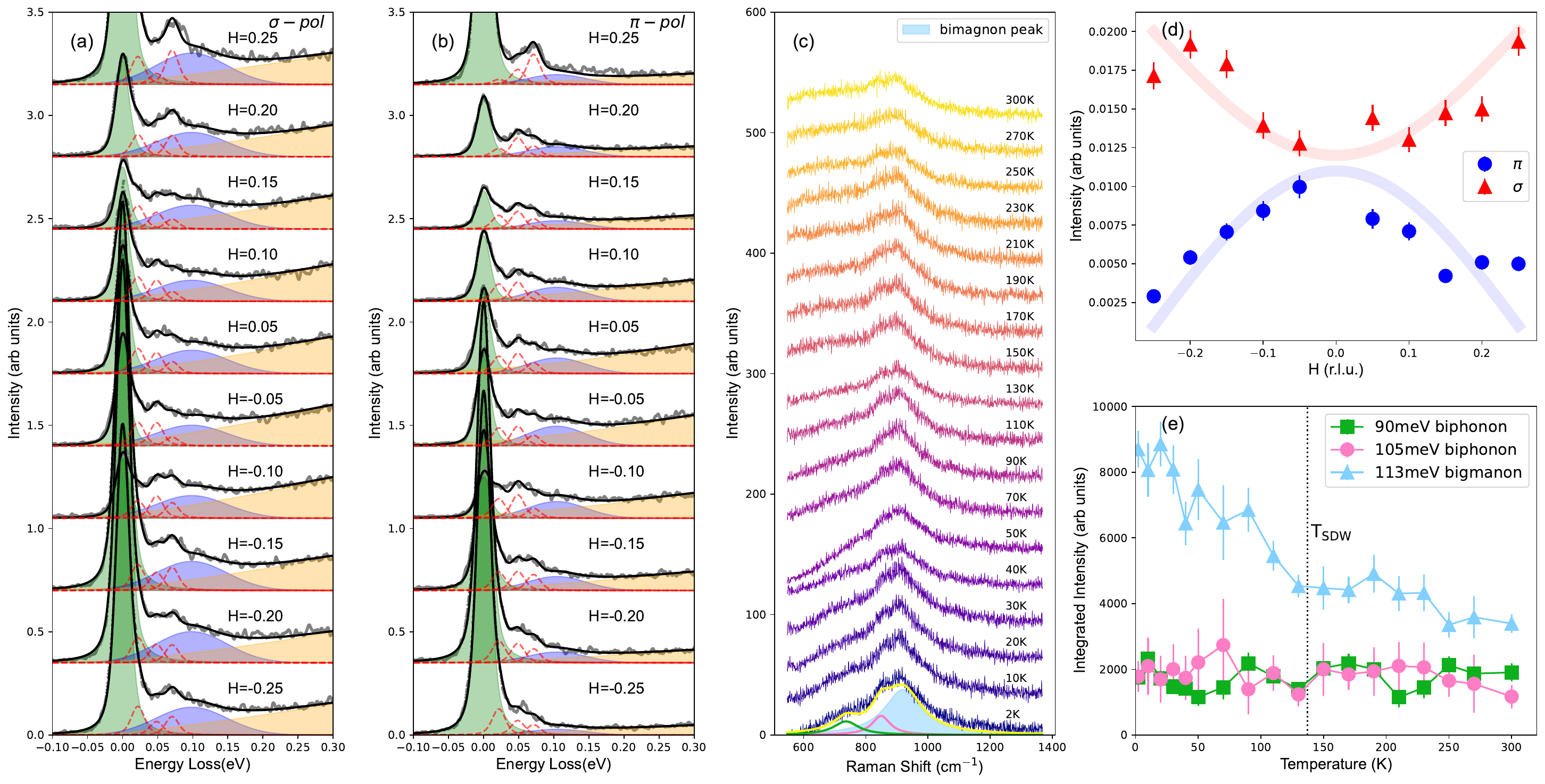}
    \caption{
        \label{fig4} 
        \textbf{(a, b)} Fitting of low-energy RIXS spectra, including elastic peak (green peak), phonons (dashed line), bimagnon (blue peak), and high-energy background (yellow tail). Each spectrum is shifted vertically for clarity.
        \textbf{(c)} Raman spectra at different temperatures, which can be fitted by three peaks, including biphonons (solid lines) and bimagons (shaded area). Each spectrum is shifted vertically for clarity. 
        \textbf{(d)} H-dependence behavior of bimagnon intensity at different polarizations extracted by fitting RIXS spectra. 
        \textbf{(e)} Temperature-dependent behavior of three Raman peaks. Biphonon do not change with temperature, whereas the bimagnon weakens with increasing temperature.
        }
\end{figure*}

For the low-energy spectra, we resolved three phonons, a broad excitation near $\sim$100 meV, and a tail from high-energy excitations (see Fig.~\ref{fig4}(a,b)); fitting details are provided in the supplementary materials. The phonon energies are consistent with previous Raman studies~\cite{Raman_india}. Here, we focus on the broad dispersionless peak at $\sim$100 meV and assign it to a bimagnon excitation for several reasons. First, its energy is approximately twice the magnon energy of La$_4$Ni$_3$O$_{10}$ observed by Ni $L$-edge RIXS~\cite{LNO4310_RIXS_PRB}. 
Second, although the absence of spin-orbit coupling in the $K$-shell orbitals prevents direct single-magnon excitation~\cite{Magnon_RIXS_PRL}, previous studies in cuprates have shown that O $K$-edge RIXS can probe bimagnons~\cite{cuprate_OK_bimagnon}. As illustrated in Fig.~\ref{fig4}(d), the intensity of this mode exhibits a pronounced polarization dependence: it increases with momentum in $\sigma$-polarization but decreases with momentum in $\pi$-polarization. Such cross-section behavior indicates that it is difficult to excite the mode with out-of-plane polarization, in sharp contrast to the phonon response (see supplementary materials) and consistent with the bimagnon behavior reported in cuprates~\cite{cuprate_OK_bimagnon}.
We also performed temperature-dependent Raman measurements in Fig.~\ref{fig4}(c). We found a dominant peak at $\sim$ 113 meV, consistent with the bimagnon observed in RIXS. Two weaker and lower-energy peaks at 90 meV and 105 meV correspond to biphonons, each linked to single-phonon modes in the low-frequency Raman spectrum (see supplementary materials). As shown in Fig.~\ref{fig4}(e), the biphonon intensities are essentially temperature independent, whereas the bimagnon peak decreases markedly with increasing temperature but persists above $T_\mathrm{SDW}$ of 137 K, indicating the presence of short-range magnetic fluctuations above the spin-density-wave transition.

Our results elucidate distinct contributions of the Ni $d_{x^2-y^2}$ and $d_{z^2}$ to electronic properties in La$_4$Ni$_3$O$_{10}$. For the out-of-plane orbitals formed by Ni $d_{z^2}$ and O $p_z$, the XAS spectrum shows a broader prepeak of O $p_z$, while EELS reveals a lower intensity and slightly reduced energy for the prepeak of outer-layer apical oxygens compared to inner-layer apical oxygens. These observations suggest that the unoccupied $d_{z^2}$ orbital consists of multiple energy states arising from chemical bonding. The absence of low-energy excitations below 1 eV from the $d_{z^2}$ orbitals further indicates its localized nature. In contrast, the in-plane orbitals formed by Ni $d_{x^2 - y^2}$ and O $p_{x,y}$ exhibit a sharper XAS prepeak and a broad charge excitation around 0.5 eV, implying that electrons in $d_{x^2 - y^2}$ can be more easily excited, and thus higher itinerancy. Our findings demonstrate that the $d_{x^2-y^2}$ and $d_{z^2}$ orbitals behave independently and possess distinct electronic behaviors.

The difficulty of exciting the bimagnon with out-of-plane polarization provides important insight into its microscopic origin. In cuprates, the bimagnon observed in O $K$-edge RIXS arises from spin exchange between two Cu $d_{x^2 - y^2}$ orbitals~\cite{Cuprate_bimagnon_Raman,cuprate_OK_bimagnon}. Considering the similar form factors between the nicklates and cuprates, one naive understanding is that the spin in nickelates also resides in the $d_{x^2 - y^2}$ orbitals. However, such a simple interpretation ignores the multilayer nature of La$_4$Ni$_3$O$_{10}$. In addition to the bonding and anti-bonding states in a bilayer nickelate, the trilayer configuration also involves non-bonding states on the outer Ni sites~\cite{LNO4310_struct_pre_WM,LNO4310_theory_nobounding_YDX_PRB2024}. 
Motivated by neutron scattering results on La$_4$Ni$_3$O$_{10}$ showing magnetism confined to the outer Ni layers~\cite{LNO4310_SDW_NC}, a more plausible scenario is that the magnetism originates from non-bonding $d_{z^2}$ orbitals at the outer layers. They can couple to $d_{x^{2}-y^{2}}$ orbitals via a strong on-site Hund's coupling $J_H$ of the order of 1eV~\cite{LNO_cal_JH}. In this framework, out-of-plane polarized excitation at an inner apical oxygen site cannot generate bimagnons, while in-plane polarization can flip two neighboring $d_{x^{2}-y^{2}}$ spins while simultaneously aligning their on-site $d_{z^{2}}$ spins. In our RIXS measurements, excitation of planar O $p_{x,y}$ orbitals mediates spin exchange between intralayer Ni atoms, producing a bimagnon whose energy reflects the underlying superexchange interactions. If magnetic interactions in La$_4$Ni$_3$O$_{10}$ is likewise dominated by interlayer coupling $J_z$, as proposed in La$_3$Ni$_2$O$_7$~\cite{LNO327_RIXS_NC,LNO327_neutron_WM}, such excitations would break two $J_z$ bonds. From the observed bimagnon energy, we estimate $SJ_z$ $\sim$ 56 meV in an effective spin model, which may be relevant for future theoretical work.

In conclusion, we establish that ligand holes reside in the planar O $p_{x,y}$ and apical O $p_z$ orbitals, hybridized with Ni $d_{x^2-y^2}$ and $d_{z^2}$, respectively. Low-energy charge excitations are dominated by the $d_{x^2-y^2}$ electrons, whereas the $d_{z^2}$ orbitals remain largely localized, with negligible excitations below 1eV. Combining RIXS and Raman spectroscopy, we identify a bimagnon at $\sim$0.1eV, which diminishes with increasing temperature but persists above T$_{SDW}$, indicating short-range spin fluctuations. This bimagnon arises from intralayer spin exchange and implies an interlayer superexchange $J_z \sim 56$~meV. Our results provide the first direct experimental insight into the orbital-specific contributions to the electronic and magnetic structures of RP-phase nickelate superconductors.

\vspace{1 ex}
\begin{acknowledgments}
\noindent
 We acknowledge enlightening discussions with Qiang-Hua Wang, Guang-Ming Zhang and Fa Wang regarding our results. Y.Y.P. is grateful for financial support from the National Natural Science Foundation of China (Grants No.12374143 and No. 11974029), the Ministry of Science and Technology of China (Grants No. 2019YFA0308401 and No. 2021YFA1401903), and Beijing Natural Science Foundation (Grant No. JQ24001). Y.L. acknowledges the financial support from the Ministry of Science and Technology of China (Grant No. 2022YFA1403000) and the National Natural Science Foundation of China (Grant No. 12274207). M.W. acknowledges the National Natural Science Foundation of China (Grant No. 12425404), the National Key Research and Development Program of China (Grant No. 2023YFA1406500), the Guangdong Basic and Applied Basic Research Funds (Grant No. 2024B1515020040), the CAS Superconducting Research Project (Grant No. SCZX-0101), the Guangdong Provincial Key Laboratory of Magnetoelectric Physics and Devices (Grant No. 2022B1212010008), and Research Center for Magnetoelectric Physics of Guangdong Province (Grant No. 2024B0303390001). Z. Chen acknowledges the financial support from the National Natural Science Foundation of China (Grant No. 52273227). 
The RIXS experimental data were collected at beamline 41A of the National Synchrotron Radiation Research Center (NSRRC) in Hsinchu 30076, Taiwan.
\end{acknowledgments}


\begin{thebibliography}{50}%
\makeatletter
\providecommand \@ifxundefined [1]{%
 \@ifx{#1\undefined}
}%
\providecommand \@ifnum [1]{%
 \ifnum #1\expandafter \@firstoftwo
 \else \expandafter \@secondoftwo
 \fi
}%
\providecommand \@ifx [1]{%
 \ifx #1\expandafter \@firstoftwo
 \else \expandafter \@secondoftwo
 \fi
}%
\providecommand \natexlab [1]{#1}%
\providecommand \enquote  [1]{``#1''}%
\providecommand \bibnamefont  [1]{#1}%
\providecommand \bibfnamefont [1]{#1}%
\providecommand \citenamefont [1]{#1}%
\providecommand \href@noop [0]{\@secondoftwo}%
\providecommand \href [0]{\begingroup \@sanitize@url \@href}%
\providecommand \@href[1]{\@@startlink{#1}\@@href}%
\providecommand \@@href[1]{\endgroup#1\@@endlink}%
\providecommand \@sanitize@url [0]{\catcode `\\12\catcode `\$12\catcode `\&12\catcode `\#12\catcode `\^12\catcode `\_12\catcode `\%12\relax}%
\providecommand \@@startlink[1]{}%
\providecommand \@@endlink[0]{}%
\providecommand \url  [0]{\begingroup\@sanitize@url \@url }%
\providecommand \@url [1]{\endgroup\@href {#1}{\urlprefix }}%
\providecommand \urlprefix  [0]{URL }%
\providecommand \Eprint [0]{\href }%
\providecommand \doibase [0]{https://doi.org/}%
\providecommand \selectlanguage [0]{\@gobble}%
\providecommand \bibinfo  [0]{\@secondoftwo}%
\providecommand \bibfield  [0]{\@secondoftwo}%
\providecommand \translation [1]{[#1]}%
\providecommand \BibitemOpen [0]{}%
\providecommand \bibitemStop [0]{}%
\providecommand \bibitemNoStop [0]{.\EOS\space}%
\providecommand \EOS [0]{\spacefactor3000\relax}%
\providecommand \BibitemShut  [1]{\csname bibitem#1\endcsname}%
\let\auto@bib@innerbib\@empty
\bibitem [{\citenamefont {Sun}\ \emph {et~al.}(2023)\citenamefont {Sun}, \citenamefont {Huo}, \citenamefont {Hu}, \citenamefont {Li}, \citenamefont {Liu}, \citenamefont {Han}, \citenamefont {Tang}, \citenamefont {Mao}, \citenamefont {Yang}, \citenamefont {Wang}, \citenamefont {Cheng}, \citenamefont {Yao}, \citenamefont {Zhang},\ and\ \citenamefont {Wang}}]{LNO327_SC_WM_Nat}%
  \BibitemOpen
  \bibfield  {author} {\bibinfo {author} {\bibfnamefont {H.}~\bibnamefont {Sun}}, \bibinfo {author} {\bibfnamefont {M.}~\bibnamefont {Huo}}, \bibinfo {author} {\bibfnamefont {X.}~\bibnamefont {Hu}}, \bibinfo {author} {\bibfnamefont {J.}~\bibnamefont {Li}}, \bibinfo {author} {\bibfnamefont {Z.}~\bibnamefont {Liu}}, \bibinfo {author} {\bibfnamefont {Y.}~\bibnamefont {Han}}, \bibinfo {author} {\bibfnamefont {L.}~\bibnamefont {Tang}}, \bibinfo {author} {\bibfnamefont {Z.}~\bibnamefont {Mao}}, \bibinfo {author} {\bibfnamefont {P.}~\bibnamefont {Yang}}, \bibinfo {author} {\bibfnamefont {B.}~\bibnamefont {Wang}}, \bibinfo {author} {\bibfnamefont {J.}~\bibnamefont {Cheng}}, \bibinfo {author} {\bibfnamefont {D.-X.}\ \bibnamefont {Yao}}, \bibinfo {author} {\bibfnamefont {G.-M.}\ \bibnamefont {Zhang}},\ and\ \bibinfo {author} {\bibfnamefont {M.}~\bibnamefont {Wang}},\ }\bibfield  {title} {\bibinfo {title} {{Signatures of superconductivity near 80 K in a nickelate under high pressure}},\ }\href
  {https://doi.org/10.1038/s41586-023-06408-7} {\bibfield  {journal} {\bibinfo  {journal} {Nature}\ }\textbf {\bibinfo {volume} {621}},\ \bibinfo {pages} {493} (\bibinfo {year} {2023})}\BibitemShut {NoStop}%
\bibitem [{\citenamefont {Wang}\ \emph {et~al.}(2024{\natexlab{a}})\citenamefont {Wang}, \citenamefont {Wang}, \citenamefont {Shen}, \citenamefont {Hou}, \citenamefont {Luo}, \citenamefont {Ma}, \citenamefont {Yang}, \citenamefont {Shi}, \citenamefont {Dou}, \citenamefont {Feng}, \citenamefont {Yang}, \citenamefont {Shi}, \citenamefont {Ren}, \citenamefont {Ma}, \citenamefont {Yang}, \citenamefont {Liu}, \citenamefont {Liu}, \citenamefont {Zhang}, \citenamefont {Dong}, \citenamefont {Wang}, \citenamefont {Jiang}, \citenamefont {Hu}, \citenamefont {Nagasaki}, \citenamefont {Kitagawa}, \citenamefont {Calder}, \citenamefont {Yan}, \citenamefont {Sun}, \citenamefont {Wang}, \citenamefont {Zhou}, \citenamefont {Uwatoko},\ and\ \citenamefont {Cheng}}]{LPNO327_SC_CJG_nat}%
  \BibitemOpen
  \bibfield  {author} {\bibinfo {author} {\bibfnamefont {N.}~\bibnamefont {Wang}}, \bibinfo {author} {\bibfnamefont {G.}~\bibnamefont {Wang}}, \bibinfo {author} {\bibfnamefont {X.}~\bibnamefont {Shen}}, \bibinfo {author} {\bibfnamefont {J.}~\bibnamefont {Hou}}, \bibinfo {author} {\bibfnamefont {J.}~\bibnamefont {Luo}}, \bibinfo {author} {\bibfnamefont {X.}~\bibnamefont {Ma}}, \bibinfo {author} {\bibfnamefont {H.}~\bibnamefont {Yang}}, \bibinfo {author} {\bibfnamefont {L.}~\bibnamefont {Shi}}, \bibinfo {author} {\bibfnamefont {J.}~\bibnamefont {Dou}}, \bibinfo {author} {\bibfnamefont {J.}~\bibnamefont {Feng}}, \bibinfo {author} {\bibfnamefont {J.}~\bibnamefont {Yang}}, \bibinfo {author} {\bibfnamefont {Y.}~\bibnamefont {Shi}}, \bibinfo {author} {\bibfnamefont {Z.}~\bibnamefont {Ren}}, \bibinfo {author} {\bibfnamefont {H.}~\bibnamefont {Ma}}, \bibinfo {author} {\bibfnamefont {P.}~\bibnamefont {Yang}}, \bibinfo {author} {\bibfnamefont {Z.}~\bibnamefont {Liu}}, \bibinfo {author} {\bibfnamefont {Y.}~\bibnamefont
  {Liu}}, \bibinfo {author} {\bibfnamefont {H.}~\bibnamefont {Zhang}}, \bibinfo {author} {\bibfnamefont {X.}~\bibnamefont {Dong}}, \bibinfo {author} {\bibfnamefont {Y.}~\bibnamefont {Wang}}, \bibinfo {author} {\bibfnamefont {K.}~\bibnamefont {Jiang}}, \bibinfo {author} {\bibfnamefont {J.}~\bibnamefont {Hu}}, \bibinfo {author} {\bibfnamefont {S.}~\bibnamefont {Nagasaki}}, \bibinfo {author} {\bibfnamefont {K.}~\bibnamefont {Kitagawa}}, \bibinfo {author} {\bibfnamefont {S.}~\bibnamefont {Calder}}, \bibinfo {author} {\bibfnamefont {J.}~\bibnamefont {Yan}}, \bibinfo {author} {\bibfnamefont {J.}~\bibnamefont {Sun}}, \bibinfo {author} {\bibfnamefont {B.}~\bibnamefont {Wang}}, \bibinfo {author} {\bibfnamefont {R.}~\bibnamefont {Zhou}}, \bibinfo {author} {\bibfnamefont {Y.}~\bibnamefont {Uwatoko}},\ and\ \bibinfo {author} {\bibfnamefont {J.}~\bibnamefont {Cheng}},\ }\bibfield  {title} {\bibinfo {title} {{Bulk high-temperature superconductivity in pressurized tetragonal La$_2$PrNi$_2$O$_7$}},\ }\href
  {https://doi.org/10.1038/s41586-024-07996-8} {\bibfield  {journal} {\bibinfo  {journal} {Nature}\ }\textbf {\bibinfo {volume} {634}},\ \bibinfo {pages} {579} (\bibinfo {year} {2024}{\natexlab{a}})}\BibitemShut {NoStop}%
\bibitem [{\citenamefont {Wang}\ \emph {et~al.}(2024{\natexlab{b}})\citenamefont {Wang}, \citenamefont {Wang}, \citenamefont {Shen}, \citenamefont {Hou}, \citenamefont {Ma}, \citenamefont {Shi}, \citenamefont {Ren}, \citenamefont {Gu}, \citenamefont {Ma}, \citenamefont {Yang}, \citenamefont {Liu}, \citenamefont {Guo}, \citenamefont {Sun}, \citenamefont {Zhang}, \citenamefont {Calder}, \citenamefont {Yan}, \citenamefont {Wang}, \citenamefont {Uwatoko},\ and\ \citenamefont {Cheng}}]{LNO327_SC_CJG_PRX}%
  \BibitemOpen
  \bibfield  {author} {\bibinfo {author} {\bibfnamefont {G.}~\bibnamefont {Wang}}, \bibinfo {author} {\bibfnamefont {N.~N.}\ \bibnamefont {Wang}}, \bibinfo {author} {\bibfnamefont {X.~L.}\ \bibnamefont {Shen}}, \bibinfo {author} {\bibfnamefont {J.}~\bibnamefont {Hou}}, \bibinfo {author} {\bibfnamefont {L.}~\bibnamefont {Ma}}, \bibinfo {author} {\bibfnamefont {L.~F.}\ \bibnamefont {Shi}}, \bibinfo {author} {\bibfnamefont {Z.~A.}\ \bibnamefont {Ren}}, \bibinfo {author} {\bibfnamefont {Y.~D.}\ \bibnamefont {Gu}}, \bibinfo {author} {\bibfnamefont {H.~M.}\ \bibnamefont {Ma}}, \bibinfo {author} {\bibfnamefont {P.~T.}\ \bibnamefont {Yang}}, \bibinfo {author} {\bibfnamefont {Z.~Y.}\ \bibnamefont {Liu}}, \bibinfo {author} {\bibfnamefont {H.~Z.}\ \bibnamefont {Guo}}, \bibinfo {author} {\bibfnamefont {J.~P.}\ \bibnamefont {Sun}}, \bibinfo {author} {\bibfnamefont {G.~M.}\ \bibnamefont {Zhang}}, \bibinfo {author} {\bibfnamefont {S.}~\bibnamefont {Calder}}, \bibinfo {author} {\bibfnamefont {J.~Q.}\ \bibnamefont {Yan}},
  \bibinfo {author} {\bibfnamefont {B.~S.}\ \bibnamefont {Wang}}, \bibinfo {author} {\bibfnamefont {Y.}~\bibnamefont {Uwatoko}},\ and\ \bibinfo {author} {\bibfnamefont {J.~G.}\ \bibnamefont {Cheng}},\ }\bibfield  {title} {\bibinfo {title} {{Pressure-Induced Superconductivity In Polycrystalline La$_3$Ni$_2$O$_{7-\delta}$}},\ }\href {https://doi.org/10.1103/PhysRevX.14.011040} {\bibfield  {journal} {\bibinfo  {journal} {Phys. Rev. X}\ }\textbf {\bibinfo {volume} {14}},\ \bibinfo {pages} {011040} (\bibinfo {year} {2024}{\natexlab{b}})}\BibitemShut {NoStop}%
\bibitem [{\citenamefont {Zhu}\ \emph {et~al.}(2024)\citenamefont {Zhu}, \citenamefont {Peng}, \citenamefont {Zhang}, \citenamefont {Pan}, \citenamefont {Chen}, \citenamefont {Chen}, \citenamefont {Ren}, \citenamefont {Liu}, \citenamefont {Hao}, \citenamefont {Li}, \citenamefont {Xing}, \citenamefont {Lan}, \citenamefont {Han}, \citenamefont {Wang}, \citenamefont {Jia}, \citenamefont {Wo}, \citenamefont {Gu}, \citenamefont {Gu}, \citenamefont {Ji}, \citenamefont {Wang}, \citenamefont {Gou}, \citenamefont {Shen}, \citenamefont {Ying}, \citenamefont {Chen}, \citenamefont {Yang}, \citenamefont {Cao}, \citenamefont {Zheng}, \citenamefont {Zeng}, \citenamefont {Guo},\ and\ \citenamefont {Zhao}}]{LNO4310_SC_ZJ_Nat}%
  \BibitemOpen
  \bibfield  {author} {\bibinfo {author} {\bibfnamefont {Y.}~\bibnamefont {Zhu}}, \bibinfo {author} {\bibfnamefont {D.}~\bibnamefont {Peng}}, \bibinfo {author} {\bibfnamefont {E.}~\bibnamefont {Zhang}}, \bibinfo {author} {\bibfnamefont {B.}~\bibnamefont {Pan}}, \bibinfo {author} {\bibfnamefont {X.}~\bibnamefont {Chen}}, \bibinfo {author} {\bibfnamefont {L.}~\bibnamefont {Chen}}, \bibinfo {author} {\bibfnamefont {H.}~\bibnamefont {Ren}}, \bibinfo {author} {\bibfnamefont {F.}~\bibnamefont {Liu}}, \bibinfo {author} {\bibfnamefont {Y.}~\bibnamefont {Hao}}, \bibinfo {author} {\bibfnamefont {N.}~\bibnamefont {Li}}, \bibinfo {author} {\bibfnamefont {Z.}~\bibnamefont {Xing}}, \bibinfo {author} {\bibfnamefont {F.}~\bibnamefont {Lan}}, \bibinfo {author} {\bibfnamefont {J.}~\bibnamefont {Han}}, \bibinfo {author} {\bibfnamefont {J.}~\bibnamefont {Wang}}, \bibinfo {author} {\bibfnamefont {D.}~\bibnamefont {Jia}}, \bibinfo {author} {\bibfnamefont {H.}~\bibnamefont {Wo}}, \bibinfo {author} {\bibfnamefont {Y.}~\bibnamefont
  {Gu}}, \bibinfo {author} {\bibfnamefont {Y.}~\bibnamefont {Gu}}, \bibinfo {author} {\bibfnamefont {L.}~\bibnamefont {Ji}}, \bibinfo {author} {\bibfnamefont {W.}~\bibnamefont {Wang}}, \bibinfo {author} {\bibfnamefont {H.}~\bibnamefont {Gou}}, \bibinfo {author} {\bibfnamefont {Y.}~\bibnamefont {Shen}}, \bibinfo {author} {\bibfnamefont {T.}~\bibnamefont {Ying}}, \bibinfo {author} {\bibfnamefont {X.}~\bibnamefont {Chen}}, \bibinfo {author} {\bibfnamefont {W.}~\bibnamefont {Yang}}, \bibinfo {author} {\bibfnamefont {H.}~\bibnamefont {Cao}}, \bibinfo {author} {\bibfnamefont {C.}~\bibnamefont {Zheng}}, \bibinfo {author} {\bibfnamefont {Q.}~\bibnamefont {Zeng}}, \bibinfo {author} {\bibfnamefont {J.-g.}\ \bibnamefont {Guo}},\ and\ \bibinfo {author} {\bibfnamefont {J.}~\bibnamefont {Zhao}},\ }\bibfield  {title} {\bibinfo {title} {{Superconductivity in pressurized trilayer La$_4$Ni$_3$O$_{10-\delta}$ single crystals}},\ }\href {https://doi.org/10.1038/s41586-024-07553-3} {\bibfield  {journal} {\bibinfo  {journal}
  {Nature}\ }\textbf {\bibinfo {volume} {631}},\ \bibinfo {pages} {531} (\bibinfo {year} {2024})}\BibitemShut {NoStop}%
\bibitem [{\citenamefont {Huang}\ \emph {et~al.}(2024)\citenamefont {Huang}, \citenamefont {Zhang}, \citenamefont {Li}, \citenamefont {Huo}, \citenamefont {Chen}, \citenamefont {Qiu}, \citenamefont {Ma}, \citenamefont {Huang}, \citenamefont {Sun},\ and\ \citenamefont {Wang}}]{PNO4310_SC_WM_CPL}%
  \BibitemOpen
  \bibfield  {author} {\bibinfo {author} {\bibfnamefont {X.}~\bibnamefont {Huang}}, \bibinfo {author} {\bibfnamefont {H.}~\bibnamefont {Zhang}}, \bibinfo {author} {\bibfnamefont {J.}~\bibnamefont {Li}}, \bibinfo {author} {\bibfnamefont {M.}~\bibnamefont {Huo}}, \bibinfo {author} {\bibfnamefont {J.}~\bibnamefont {Chen}}, \bibinfo {author} {\bibfnamefont {Z.}~\bibnamefont {Qiu}}, \bibinfo {author} {\bibfnamefont {P.}~\bibnamefont {Ma}}, \bibinfo {author} {\bibfnamefont {C.}~\bibnamefont {Huang}}, \bibinfo {author} {\bibfnamefont {H.}~\bibnamefont {Sun}},\ and\ \bibinfo {author} {\bibfnamefont {M.}~\bibnamefont {Wang}},\ }\bibfield  {title} {\bibinfo {title} {{Signature of Superconductivity in Pressurized Trilayer-Nickelate Pr$_4$Ni$_3$O$_{10-\delta}$}},\ }\bibfield  {journal} {\bibinfo  {journal} {Chinese Physics Letters}\ }\textbf {\bibinfo {volume} {41}},\ \href {https://doi.org/10.1088/0256-307x/41/12/127403} {10.1088/0256-307x/41/12/127403} (\bibinfo {year} {2024})\BibitemShut {NoStop}%
\bibitem [{\citenamefont {Li}\ \emph {et~al.}(2024{\natexlab{a}})\citenamefont {Li}, \citenamefont {Zhang}, \citenamefont {Xiang}, \citenamefont {Zhang}, \citenamefont {Zhu},\ and\ \citenamefont {Wen}}]{LNO4310_SC_WHH_CPL}%
  \BibitemOpen
  \bibfield  {author} {\bibinfo {author} {\bibfnamefont {Q.}~\bibnamefont {Li}}, \bibinfo {author} {\bibfnamefont {Y.-J.}\ \bibnamefont {Zhang}}, \bibinfo {author} {\bibfnamefont {Z.-N.}\ \bibnamefont {Xiang}}, \bibinfo {author} {\bibfnamefont {Y.}~\bibnamefont {Zhang}}, \bibinfo {author} {\bibfnamefont {X.}~\bibnamefont {Zhu}},\ and\ \bibinfo {author} {\bibfnamefont {H.-H.}\ \bibnamefont {Wen}},\ }\bibfield  {title} {\bibinfo {title} {Signature of superconductivity in pressurized la$_4$ni$_3$o$_{10}$},\ }\bibfield  {journal} {\bibinfo  {journal} {Chinese Physics Letters}\ }\textbf {\bibinfo {volume} {41}},\ \href {https://doi.org/10.1088/0256-307x/41/1/017401} {10.1088/0256-307x/41/1/017401} (\bibinfo {year} {2024}{\natexlab{a}})\BibitemShut {NoStop}%
\bibitem [{\citenamefont {Nagata}\ \emph {et~al.}(2024)\citenamefont {Nagata}, \citenamefont {Sakurai}, \citenamefont {Ueki}, \citenamefont {Yamane}, \citenamefont {Matsumoto}, \citenamefont {Terashima}, \citenamefont {Hirose}, \citenamefont {Ohta}, \citenamefont {Kato},\ and\ \citenamefont {Takano}}]{LNO4310_SC_doping_Japan}%
  \BibitemOpen
  \bibfield  {author} {\bibinfo {author} {\bibfnamefont {H.}~\bibnamefont {Nagata}}, \bibinfo {author} {\bibfnamefont {H.}~\bibnamefont {Sakurai}}, \bibinfo {author} {\bibfnamefont {Y.}~\bibnamefont {Ueki}}, \bibinfo {author} {\bibfnamefont {K.}~\bibnamefont {Yamane}}, \bibinfo {author} {\bibfnamefont {R.}~\bibnamefont {Matsumoto}}, \bibinfo {author} {\bibfnamefont {K.}~\bibnamefont {Terashima}}, \bibinfo {author} {\bibfnamefont {K.}~\bibnamefont {Hirose}}, \bibinfo {author} {\bibfnamefont {H.}~\bibnamefont {Ohta}}, \bibinfo {author} {\bibfnamefont {M.}~\bibnamefont {Kato}},\ and\ \bibinfo {author} {\bibfnamefont {Y.}~\bibnamefont {Takano}},\ }\bibfield  {title} {\bibinfo {title} {{Pressure-Induced Superconductivity in La$_4$Ni$_3$O$_{10+\delta}$ ($\delta$ = 0.04 and -0.01)}},\ }\bibfield  {journal} {\bibinfo  {journal} {Journal of the Physical Society of Japan}\ }\textbf {\bibinfo {volume} {93}},\ \href {https://doi.org/10.7566/jpsj.93.095003} {10.7566/jpsj.93.095003} (\bibinfo {year} {2024})\BibitemShut
  {NoStop}%
\bibitem [{\citenamefont {Ko}\ \emph {et~al.}(2024)\citenamefont {Ko}, \citenamefont {Yu}, \citenamefont {Liu}, \citenamefont {Bhatt}, \citenamefont {Li}, \citenamefont {Thampy}, \citenamefont {Kuo}, \citenamefont {Wang}, \citenamefont {Lee}, \citenamefont {Lee}, \citenamefont {Lee}, \citenamefont {Goodge}, \citenamefont {Muller},\ and\ \citenamefont {Hwang}}]{LNO327film_SC_Nat}%
  \BibitemOpen
  \bibfield  {author} {\bibinfo {author} {\bibfnamefont {E.~K.}\ \bibnamefont {Ko}}, \bibinfo {author} {\bibfnamefont {Y.}~\bibnamefont {Yu}}, \bibinfo {author} {\bibfnamefont {Y.}~\bibnamefont {Liu}}, \bibinfo {author} {\bibfnamefont {L.}~\bibnamefont {Bhatt}}, \bibinfo {author} {\bibfnamefont {J.}~\bibnamefont {Li}}, \bibinfo {author} {\bibfnamefont {V.}~\bibnamefont {Thampy}}, \bibinfo {author} {\bibfnamefont {C.-T.}\ \bibnamefont {Kuo}}, \bibinfo {author} {\bibfnamefont {B.~Y.}\ \bibnamefont {Wang}}, \bibinfo {author} {\bibfnamefont {Y.}~\bibnamefont {Lee}}, \bibinfo {author} {\bibfnamefont {K.}~\bibnamefont {Lee}}, \bibinfo {author} {\bibfnamefont {J.-S.}\ \bibnamefont {Lee}}, \bibinfo {author} {\bibfnamefont {B.~H.}\ \bibnamefont {Goodge}}, \bibinfo {author} {\bibfnamefont {D.~A.}\ \bibnamefont {Muller}},\ and\ \bibinfo {author} {\bibfnamefont {H.~Y.}\ \bibnamefont {Hwang}},\ }\bibfield  {title} {\bibinfo {title} {{Signatures of ambient pressure superconductivity in thin film La$_3$Ni$_2$O$_{7}$}},\
  }\bibfield  {journal} {\bibinfo  {journal} {Nature}\ }\href {https://doi.org/10.1038/s41586-024-08525-3} {10.1038/s41586-024-08525-3} (\bibinfo {year} {2024})\BibitemShut {NoStop}%
\bibitem [{\citenamefont {Zhou}\ \emph {et~al.}(2025)\citenamefont {Zhou}, \citenamefont {Lv}, \citenamefont {Wang}, \citenamefont {Nie}, \citenamefont {Chen}, \citenamefont {Li}, \citenamefont {Huang}, \citenamefont {Chen}, \citenamefont {Sun}, \citenamefont {Xue},\ and\ \citenamefont {Chen}}]{LPNOfilm_SC_XQK_Nat}%
  \BibitemOpen
  \bibfield  {author} {\bibinfo {author} {\bibfnamefont {G.}~\bibnamefont {Zhou}}, \bibinfo {author} {\bibfnamefont {W.}~\bibnamefont {Lv}}, \bibinfo {author} {\bibfnamefont {H.}~\bibnamefont {Wang}}, \bibinfo {author} {\bibfnamefont {Z.}~\bibnamefont {Nie}}, \bibinfo {author} {\bibfnamefont {Y.}~\bibnamefont {Chen}}, \bibinfo {author} {\bibfnamefont {Y.}~\bibnamefont {Li}}, \bibinfo {author} {\bibfnamefont {H.}~\bibnamefont {Huang}}, \bibinfo {author} {\bibfnamefont {W.-Q.}\ \bibnamefont {Chen}}, \bibinfo {author} {\bibfnamefont {Y.-J.}\ \bibnamefont {Sun}}, \bibinfo {author} {\bibfnamefont {Q.-K.}\ \bibnamefont {Xue}},\ and\ \bibinfo {author} {\bibfnamefont {Z.}~\bibnamefont {Chen}},\ }\bibfield  {title} {\bibinfo {title} {{Ambient-pressure superconductivity onset above 40 K in (La,Pr)$_3$Ni$_2$O$_7$ films}},\ }\href {https://doi.org/10.1038/s41586-025-08755-z} {\bibfield  {journal} {\bibinfo  {journal} {Nature}\ }\textbf {\bibinfo {volume} {640}},\ \bibinfo {pages} {641} (\bibinfo {year}
  {2025})}\BibitemShut {NoStop}%
\bibitem [{\citenamefont {Li}\ \emph {et~al.}(2019)\citenamefont {Li}, \citenamefont {Lee}, \citenamefont {Wang}, \citenamefont {Osada}, \citenamefont {Crossley}, \citenamefont {Lee}, \citenamefont {Cui}, \citenamefont {Hikita},\ and\ \citenamefont {Hwang}}]{RNO112_SC_Nat}%
  \BibitemOpen
  \bibfield  {author} {\bibinfo {author} {\bibfnamefont {D.}~\bibnamefont {Li}}, \bibinfo {author} {\bibfnamefont {K.}~\bibnamefont {Lee}}, \bibinfo {author} {\bibfnamefont {B.~Y.}\ \bibnamefont {Wang}}, \bibinfo {author} {\bibfnamefont {M.}~\bibnamefont {Osada}}, \bibinfo {author} {\bibfnamefont {S.}~\bibnamefont {Crossley}}, \bibinfo {author} {\bibfnamefont {H.~R.}\ \bibnamefont {Lee}}, \bibinfo {author} {\bibfnamefont {Y.}~\bibnamefont {Cui}}, \bibinfo {author} {\bibfnamefont {Y.}~\bibnamefont {Hikita}},\ and\ \bibinfo {author} {\bibfnamefont {H.~Y.}\ \bibnamefont {Hwang}},\ }\bibfield  {title} {\bibinfo {title} {{Superconductivity in an infinite-layer nickelate}},\ }\href {https://doi.org/10.1038/s41586-019-1496-5} {\bibfield  {journal} {\bibinfo  {journal} {Nature}\ }\textbf {\bibinfo {volume} {572}},\ \bibinfo {pages} {624} (\bibinfo {year} {2019})}\BibitemShut {NoStop}%
\bibitem [{\citenamefont {Lee}\ \emph {et~al.}(2023)\citenamefont {Lee}, \citenamefont {Wang}, \citenamefont {Osada}, \citenamefont {Goodge}, \citenamefont {Wang}, \citenamefont {Lee}, \citenamefont {Harvey}, \citenamefont {Kim}, \citenamefont {Yu}, \citenamefont {Murthy} \emph {et~al.}}]{RNO112_SC_linearR_nat}%
  \BibitemOpen
  \bibfield  {author} {\bibinfo {author} {\bibfnamefont {K.}~\bibnamefont {Lee}}, \bibinfo {author} {\bibfnamefont {B.~Y.}\ \bibnamefont {Wang}}, \bibinfo {author} {\bibfnamefont {M.}~\bibnamefont {Osada}}, \bibinfo {author} {\bibfnamefont {B.~H.}\ \bibnamefont {Goodge}}, \bibinfo {author} {\bibfnamefont {T.~C.}\ \bibnamefont {Wang}}, \bibinfo {author} {\bibfnamefont {Y.}~\bibnamefont {Lee}}, \bibinfo {author} {\bibfnamefont {S.}~\bibnamefont {Harvey}}, \bibinfo {author} {\bibfnamefont {W.~J.}\ \bibnamefont {Kim}}, \bibinfo {author} {\bibfnamefont {Y.}~\bibnamefont {Yu}}, \bibinfo {author} {\bibfnamefont {C.}~\bibnamefont {Murthy}}, \emph {et~al.},\ }\bibfield  {title} {\bibinfo {title} {{Linear-in-temperature resistivity for optimally superconducting (Nd, Sr) NiO$_2$}},\ }\href@noop {} {\bibfield  {journal} {\bibinfo  {journal} {Nature}\ }\textbf {\bibinfo {volume} {619}},\ \bibinfo {pages} {288} (\bibinfo {year} {2023})}\BibitemShut {NoStop}%
\bibitem [{\citenamefont {Parzyck}\ \emph {et~al.}(2025)\citenamefont {Parzyck}, \citenamefont {Wu}, \citenamefont {Bhatt}, \citenamefont {Kang}, \citenamefont {Arthur}, \citenamefont {Pedersen}, \citenamefont {Sutarto}, \citenamefont {Fan}, \citenamefont {Pelliciari}, \citenamefont {Bisogni}, \citenamefont {Herranz}, \citenamefont {Georgescu}, \citenamefont {Hawthorn}, \citenamefont {Kourkoutis}, \citenamefont {Muller}, \citenamefont {Schlom},\ and\ \citenamefont {Shen}}]{RNO112_SC_PRX2015}%
  \BibitemOpen
  \bibfield  {author} {\bibinfo {author} {\bibfnamefont {C.~T.}\ \bibnamefont {Parzyck}}, \bibinfo {author} {\bibfnamefont {Y.}~\bibnamefont {Wu}}, \bibinfo {author} {\bibfnamefont {L.}~\bibnamefont {Bhatt}}, \bibinfo {author} {\bibfnamefont {M.}~\bibnamefont {Kang}}, \bibinfo {author} {\bibfnamefont {Z.}~\bibnamefont {Arthur}}, \bibinfo {author} {\bibfnamefont {T.~M.}\ \bibnamefont {Pedersen}}, \bibinfo {author} {\bibfnamefont {R.}~\bibnamefont {Sutarto}}, \bibinfo {author} {\bibfnamefont {S.}~\bibnamefont {Fan}}, \bibinfo {author} {\bibfnamefont {J.}~\bibnamefont {Pelliciari}}, \bibinfo {author} {\bibfnamefont {V.}~\bibnamefont {Bisogni}}, \bibinfo {author} {\bibfnamefont {G.}~\bibnamefont {Herranz}}, \bibinfo {author} {\bibfnamefont {A.~B.}\ \bibnamefont {Georgescu}}, \bibinfo {author} {\bibfnamefont {D.~G.}\ \bibnamefont {Hawthorn}}, \bibinfo {author} {\bibfnamefont {L.~F.}\ \bibnamefont {Kourkoutis}}, \bibinfo {author} {\bibfnamefont {D.~A.}\ \bibnamefont {Muller}}, \bibinfo {author} {\bibfnamefont
  {D.~G.}\ \bibnamefont {Schlom}},\ and\ \bibinfo {author} {\bibfnamefont {K.~M.}\ \bibnamefont {Shen}},\ }\bibfield  {title} {\bibinfo {title} {{Superconductivity in the Parent Infinite-Layer Nickelate NdNiO$_{2}$}},\ }\href {https://doi.org/10.1103/PhysRevX.15.021048} {\bibfield  {journal} {\bibinfo  {journal} {Phys. Rev. X}\ }\textbf {\bibinfo {volume} {15}},\ \bibinfo {pages} {021048} (\bibinfo {year} {2025})}\BibitemShut {NoStop}%
\bibitem [{\citenamefont {Zeng}\ \emph {et~al.}(2020)\citenamefont {Zeng}, \citenamefont {Tang}, \citenamefont {Yin}, \citenamefont {Li}, \citenamefont {Li}, \citenamefont {Huang}, \citenamefont {Hu}, \citenamefont {Liu}, \citenamefont {Omar}, \citenamefont {Jani}, \citenamefont {Lim}, \citenamefont {Han}, \citenamefont {Wan}, \citenamefont {Yang}, \citenamefont {Pennycook}, \citenamefont {Wee},\ and\ \citenamefont {Ariando}}]{RNO112_SC_doping_PRL}%
  \BibitemOpen
  \bibfield  {author} {\bibinfo {author} {\bibfnamefont {S.}~\bibnamefont {Zeng}}, \bibinfo {author} {\bibfnamefont {C.~S.}\ \bibnamefont {Tang}}, \bibinfo {author} {\bibfnamefont {X.}~\bibnamefont {Yin}}, \bibinfo {author} {\bibfnamefont {C.}~\bibnamefont {Li}}, \bibinfo {author} {\bibfnamefont {M.}~\bibnamefont {Li}}, \bibinfo {author} {\bibfnamefont {Z.}~\bibnamefont {Huang}}, \bibinfo {author} {\bibfnamefont {J.}~\bibnamefont {Hu}}, \bibinfo {author} {\bibfnamefont {W.}~\bibnamefont {Liu}}, \bibinfo {author} {\bibfnamefont {G.~J.}\ \bibnamefont {Omar}}, \bibinfo {author} {\bibfnamefont {H.}~\bibnamefont {Jani}}, \bibinfo {author} {\bibfnamefont {Z.~S.}\ \bibnamefont {Lim}}, \bibinfo {author} {\bibfnamefont {K.}~\bibnamefont {Han}}, \bibinfo {author} {\bibfnamefont {D.}~\bibnamefont {Wan}}, \bibinfo {author} {\bibfnamefont {P.}~\bibnamefont {Yang}}, \bibinfo {author} {\bibfnamefont {S.~J.}\ \bibnamefont {Pennycook}}, \bibinfo {author} {\bibfnamefont {A.~T.~S.}\ \bibnamefont {Wee}},\ and\ \bibinfo {author}
  {\bibfnamefont {A.}~\bibnamefont {Ariando}},\ }\bibfield  {title} {\bibinfo {title} {{Phase Diagram and Superconducting Dome of Infinite-Layer $Nd_{1-x}Sr_xNiO_2$ Thin Films}},\ }\href {https://doi.org/10.1103/PhysRevLett.125.147003} {\bibfield  {journal} {\bibinfo  {journal} {Phys Rev Lett}\ }\textbf {\bibinfo {volume} {125}},\ \bibinfo {pages} {147003} (\bibinfo {year} {2020})}\BibitemShut {NoStop}%
\bibitem [{\citenamefont {Yang}\ \emph {et~al.}(2024{\natexlab{a}})\citenamefont {Yang}, \citenamefont {Sun}, \citenamefont {Hu}, \citenamefont {Xie}, \citenamefont {Miao}, \citenamefont {Luo}, \citenamefont {Chen}, \citenamefont {Liang}, \citenamefont {Zhu}, \citenamefont {Qu}, \citenamefont {Chen}, \citenamefont {Huo}, \citenamefont {Huang}, \citenamefont {Zhang}, \citenamefont {Zhang}, \citenamefont {Yang}, \citenamefont {Wang}, \citenamefont {Peng}, \citenamefont {Mao}, \citenamefont {Liu}, \citenamefont {Xu}, \citenamefont {Qian}, \citenamefont {Yao}, \citenamefont {Wang}, \citenamefont {Zhao},\ and\ \citenamefont {Zhou}}]{LNO327_band_ZXJ_NC}%
  \BibitemOpen
  \bibfield  {author} {\bibinfo {author} {\bibfnamefont {J.}~\bibnamefont {Yang}}, \bibinfo {author} {\bibfnamefont {H.}~\bibnamefont {Sun}}, \bibinfo {author} {\bibfnamefont {X.}~\bibnamefont {Hu}}, \bibinfo {author} {\bibfnamefont {Y.}~\bibnamefont {Xie}}, \bibinfo {author} {\bibfnamefont {T.}~\bibnamefont {Miao}}, \bibinfo {author} {\bibfnamefont {H.}~\bibnamefont {Luo}}, \bibinfo {author} {\bibfnamefont {H.}~\bibnamefont {Chen}}, \bibinfo {author} {\bibfnamefont {B.}~\bibnamefont {Liang}}, \bibinfo {author} {\bibfnamefont {W.}~\bibnamefont {Zhu}}, \bibinfo {author} {\bibfnamefont {G.}~\bibnamefont {Qu}}, \bibinfo {author} {\bibfnamefont {C.-Q.}\ \bibnamefont {Chen}}, \bibinfo {author} {\bibfnamefont {M.}~\bibnamefont {Huo}}, \bibinfo {author} {\bibfnamefont {Y.}~\bibnamefont {Huang}}, \bibinfo {author} {\bibfnamefont {S.}~\bibnamefont {Zhang}}, \bibinfo {author} {\bibfnamefont {F.}~\bibnamefont {Zhang}}, \bibinfo {author} {\bibfnamefont {F.}~\bibnamefont {Yang}}, \bibinfo {author} {\bibfnamefont
  {Z.}~\bibnamefont {Wang}}, \bibinfo {author} {\bibfnamefont {Q.}~\bibnamefont {Peng}}, \bibinfo {author} {\bibfnamefont {H.}~\bibnamefont {Mao}}, \bibinfo {author} {\bibfnamefont {G.}~\bibnamefont {Liu}}, \bibinfo {author} {\bibfnamefont {Z.}~\bibnamefont {Xu}}, \bibinfo {author} {\bibfnamefont {T.}~\bibnamefont {Qian}}, \bibinfo {author} {\bibfnamefont {D.-X.}\ \bibnamefont {Yao}}, \bibinfo {author} {\bibfnamefont {M.}~\bibnamefont {Wang}}, \bibinfo {author} {\bibfnamefont {L.}~\bibnamefont {Zhao}},\ and\ \bibinfo {author} {\bibfnamefont {X.~J.}\ \bibnamefont {Zhou}},\ }\bibfield  {title} {\bibinfo {title} {{Orbital-dependent electron correlation in double-layer nickelate La$_3$Ni$_2$O$_7$}},\ }\bibfield  {journal} {\bibinfo  {journal} {Nature Communications}\ }\textbf {\bibinfo {volume} {15}},\ \href {https://doi.org/10.1038/s41467-024-48701-7} {10.1038/s41467-024-48701-7} (\bibinfo {year} {2024}{\natexlab{a}})\BibitemShut {NoStop}%
\bibitem [{\citenamefont {Li}\ \emph {et~al.}(2017)\citenamefont {Li}, \citenamefont {Zhou}, \citenamefont {Nummy}, \citenamefont {Zhang}, \citenamefont {Pardo}, \citenamefont {Pickett}, \citenamefont {Mitchell},\ and\ \citenamefont {Dessau}}]{LNO4310_ARPES_band_NC}%
  \BibitemOpen
  \bibfield  {author} {\bibinfo {author} {\bibfnamefont {H.}~\bibnamefont {Li}}, \bibinfo {author} {\bibfnamefont {X.}~\bibnamefont {Zhou}}, \bibinfo {author} {\bibfnamefont {T.}~\bibnamefont {Nummy}}, \bibinfo {author} {\bibfnamefont {J.}~\bibnamefont {Zhang}}, \bibinfo {author} {\bibfnamefont {V.}~\bibnamefont {Pardo}}, \bibinfo {author} {\bibfnamefont {W.~E.}\ \bibnamefont {Pickett}}, \bibinfo {author} {\bibfnamefont {J.~F.}\ \bibnamefont {Mitchell}},\ and\ \bibinfo {author} {\bibfnamefont {D.~S.}\ \bibnamefont {Dessau}},\ }\bibfield  {title} {\bibinfo {title} {{Fermiology and electron dynamics of trilayer nickelate La$_4$Ni$_3$O$_{10}$}},\ }\href {https://doi.org/10.1038/s41467-017-00777-0} {\bibfield  {journal} {\bibinfo  {journal} {Nat Commun}\ }\textbf {\bibinfo {volume} {8}},\ \bibinfo {pages} {704} (\bibinfo {year} {2017})}\BibitemShut {NoStop}%
\bibitem [{\citenamefont {Lechermann}\ \emph {et~al.}(2023)\citenamefont {Lechermann}, \citenamefont {Gondolf}, \citenamefont {B\"otzel},\ and\ \citenamefont {Eremin}}]{LNO327_Theory_band_PRB2023}%
  \BibitemOpen
  \bibfield  {author} {\bibinfo {author} {\bibfnamefont {F.}~\bibnamefont {Lechermann}}, \bibinfo {author} {\bibfnamefont {J.}~\bibnamefont {Gondolf}}, \bibinfo {author} {\bibfnamefont {S.}~\bibnamefont {B\"otzel}},\ and\ \bibinfo {author} {\bibfnamefont {I.~M.}\ \bibnamefont {Eremin}},\ }\bibfield  {title} {\bibinfo {title} {{Electronic correlations and superconducting instability in ${\mathrm{La}}_{3}{\mathrm{Ni}}_{2}{\mathrm{O}}_{7}$ under high pressure}},\ }\href {https://doi.org/10.1103/PhysRevB.108.L201121} {\bibfield  {journal} {\bibinfo  {journal} {Phys. Rev. B}\ }\textbf {\bibinfo {volume} {108}},\ \bibinfo {pages} {L201121} (\bibinfo {year} {2023})}\BibitemShut {NoStop}%
\bibitem [{\citenamefont {Chen}\ \emph {et~al.}(2024{\natexlab{a}})\citenamefont {Chen}, \citenamefont {Choi}, \citenamefont {Jiang}, \citenamefont {Mei}, \citenamefont {Jiang}, \citenamefont {Li}, \citenamefont {Agrestini}, \citenamefont {Garcia-Fernandez}, \citenamefont {Sun}, \citenamefont {Huang}, \citenamefont {Shen}, \citenamefont {Wang}, \citenamefont {Hu}, \citenamefont {Lu}, \citenamefont {Zhou},\ and\ \citenamefont {Feng}}]{LNO327_RIXS_NC}%
  \BibitemOpen
  \bibfield  {author} {\bibinfo {author} {\bibfnamefont {X.}~\bibnamefont {Chen}}, \bibinfo {author} {\bibfnamefont {J.}~\bibnamefont {Choi}}, \bibinfo {author} {\bibfnamefont {Z.}~\bibnamefont {Jiang}}, \bibinfo {author} {\bibfnamefont {J.}~\bibnamefont {Mei}}, \bibinfo {author} {\bibfnamefont {K.}~\bibnamefont {Jiang}}, \bibinfo {author} {\bibfnamefont {J.}~\bibnamefont {Li}}, \bibinfo {author} {\bibfnamefont {S.}~\bibnamefont {Agrestini}}, \bibinfo {author} {\bibfnamefont {M.}~\bibnamefont {Garcia-Fernandez}}, \bibinfo {author} {\bibfnamefont {H.}~\bibnamefont {Sun}}, \bibinfo {author} {\bibfnamefont {X.}~\bibnamefont {Huang}}, \bibinfo {author} {\bibfnamefont {D.}~\bibnamefont {Shen}}, \bibinfo {author} {\bibfnamefont {M.}~\bibnamefont {Wang}}, \bibinfo {author} {\bibfnamefont {J.}~\bibnamefont {Hu}}, \bibinfo {author} {\bibfnamefont {Y.}~\bibnamefont {Lu}}, \bibinfo {author} {\bibfnamefont {K.-J.}\ \bibnamefont {Zhou}},\ and\ \bibinfo {author} {\bibfnamefont {D.}~\bibnamefont {Feng}},\ }\bibfield
  {title} {\bibinfo {title} {{Electronic and magnetic excitations in $La_3Ni_2O_7$}},\ }\bibfield  {journal} {\bibinfo  {journal} {Nature Communications}\ }\textbf {\bibinfo {volume} {15}},\ \href {https://doi.org/10.1038/s41467-024-53863-5} {10.1038/s41467-024-53863-5} (\bibinfo {year} {2024}{\natexlab{a}})\BibitemShut {NoStop}%
\bibitem [{\citenamefont {Zhang}\ \emph {et~al.}(2020)\citenamefont {Zhang}, \citenamefont {Phelan}, \citenamefont {Botana}, \citenamefont {Chen}, \citenamefont {Zheng}, \citenamefont {Krogstad}, \citenamefont {Wang}, \citenamefont {Qiu}, \citenamefont {Rodriguez-Rivera}, \citenamefont {Osborn}, \citenamefont {Rosenkranz}, \citenamefont {Norman},\ and\ \citenamefont {Mitchell}}]{LNO4310_SDW_NC}%
  \BibitemOpen
  \bibfield  {author} {\bibinfo {author} {\bibfnamefont {J.}~\bibnamefont {Zhang}}, \bibinfo {author} {\bibfnamefont {D.}~\bibnamefont {Phelan}}, \bibinfo {author} {\bibfnamefont {A.~S.}\ \bibnamefont {Botana}}, \bibinfo {author} {\bibfnamefont {Y.~S.}\ \bibnamefont {Chen}}, \bibinfo {author} {\bibfnamefont {H.}~\bibnamefont {Zheng}}, \bibinfo {author} {\bibfnamefont {M.}~\bibnamefont {Krogstad}}, \bibinfo {author} {\bibfnamefont {S.~G.}\ \bibnamefont {Wang}}, \bibinfo {author} {\bibfnamefont {Y.}~\bibnamefont {Qiu}}, \bibinfo {author} {\bibfnamefont {J.~A.}\ \bibnamefont {Rodriguez-Rivera}}, \bibinfo {author} {\bibfnamefont {R.}~\bibnamefont {Osborn}}, \bibinfo {author} {\bibfnamefont {S.}~\bibnamefont {Rosenkranz}}, \bibinfo {author} {\bibfnamefont {M.~R.}\ \bibnamefont {Norman}},\ and\ \bibinfo {author} {\bibfnamefont {J.~F.}\ \bibnamefont {Mitchell}},\ }\bibfield  {title} {\bibinfo {title} {{Intertwined density waves in a metallic nickelate}},\ }\href {https://doi.org/10.1038/s41467-020-19836-0}
  {\bibfield  {journal} {\bibinfo  {journal} {Nat Commun}\ }\textbf {\bibinfo {volume} {11}},\ \bibinfo {pages} {6003} (\bibinfo {year} {2020})}\BibitemShut {NoStop}%
\bibitem [{\citenamefont {Kakoi}\ \emph {et~al.}(2024)\citenamefont {Kakoi}, \citenamefont {Oi}, \citenamefont {Ohshita}, \citenamefont {Yashima}, \citenamefont {Kuroki}, \citenamefont {Kato}, \citenamefont {Takahashi}, \citenamefont {Ishiwata}, \citenamefont {Adachi}, \citenamefont {Hatada}, \citenamefont {Uda},\ and\ \citenamefont {Mukuda}}]{LNO_NMR_SDW}%
  \BibitemOpen
  \bibfield  {author} {\bibinfo {author} {\bibfnamefont {M.}~\bibnamefont {Kakoi}}, \bibinfo {author} {\bibfnamefont {T.}~\bibnamefont {Oi}}, \bibinfo {author} {\bibfnamefont {Y.}~\bibnamefont {Ohshita}}, \bibinfo {author} {\bibfnamefont {M.}~\bibnamefont {Yashima}}, \bibinfo {author} {\bibfnamefont {K.}~\bibnamefont {Kuroki}}, \bibinfo {author} {\bibfnamefont {T.}~\bibnamefont {Kato}}, \bibinfo {author} {\bibfnamefont {H.}~\bibnamefont {Takahashi}}, \bibinfo {author} {\bibfnamefont {S.}~\bibnamefont {Ishiwata}}, \bibinfo {author} {\bibfnamefont {Y.}~\bibnamefont {Adachi}}, \bibinfo {author} {\bibfnamefont {N.}~\bibnamefont {Hatada}}, \bibinfo {author} {\bibfnamefont {T.}~\bibnamefont {Uda}},\ and\ \bibinfo {author} {\bibfnamefont {H.}~\bibnamefont {Mukuda}},\ }\bibfield  {title} {\bibinfo {title} {{Multiband Metallic Ground State in Multilayered Nickelates La$_3$Ni$_2$O$_7$ and La$_4$Ni$_3$O$_{10}$ Probed by 139La-NMR at Ambient Pressure}},\ }\href {https://doi.org/10.7566/JPSJ.93.053702} {\bibfield
  {journal} {\bibinfo  {journal} {Journal of the Physical Society of Japan}\ }\textbf {\bibinfo {volume} {93}},\ \bibinfo {pages} {053702} (\bibinfo {year} {2024})},\ \Eprint {https://arxiv.org/abs/https://doi.org/10.7566/JPSJ.93.053702} {https://doi.org/10.7566/JPSJ.93.053702} \BibitemShut {NoStop}%
\bibitem [{\citenamefont {Meng}\ \emph {et~al.}(2024)\citenamefont {Meng}, \citenamefont {Yang}, \citenamefont {Sun}, \citenamefont {Zhang}, \citenamefont {Luo}, \citenamefont {Chen}, \citenamefont {Ma}, \citenamefont {Wang}, \citenamefont {Hong}, \citenamefont {Wang},\ and\ \citenamefont {Yu}}]{LNO327_SDW_FastR_presure_NC}%
  \BibitemOpen
  \bibfield  {author} {\bibinfo {author} {\bibfnamefont {Y.}~\bibnamefont {Meng}}, \bibinfo {author} {\bibfnamefont {Y.}~\bibnamefont {Yang}}, \bibinfo {author} {\bibfnamefont {H.}~\bibnamefont {Sun}}, \bibinfo {author} {\bibfnamefont {S.}~\bibnamefont {Zhang}}, \bibinfo {author} {\bibfnamefont {J.}~\bibnamefont {Luo}}, \bibinfo {author} {\bibfnamefont {L.}~\bibnamefont {Chen}}, \bibinfo {author} {\bibfnamefont {X.}~\bibnamefont {Ma}}, \bibinfo {author} {\bibfnamefont {M.}~\bibnamefont {Wang}}, \bibinfo {author} {\bibfnamefont {F.}~\bibnamefont {Hong}}, \bibinfo {author} {\bibfnamefont {X.}~\bibnamefont {Wang}},\ and\ \bibinfo {author} {\bibfnamefont {X.}~\bibnamefont {Yu}},\ }\bibfield  {title} {\bibinfo {title} {{Density-wave-like gap evolution in La$_3$Ni$_2$O$_7$ under high pressure revealed by ultrafast optical spectroscopy}},\ }\bibfield  {journal} {\bibinfo  {journal} {Nature Communications}\ }\textbf {\bibinfo {volume} {15}},\ \href {https://doi.org/10.1038/s41467-024-54518-1}
  {10.1038/s41467-024-54518-1} (\bibinfo {year} {2024})\BibitemShut {NoStop}%
\bibitem [{\citenamefont {Liu}\ \emph {et~al.}(2023{\natexlab{a}})\citenamefont {Liu}, \citenamefont {Sun}, \citenamefont {Huo}, \citenamefont {Ma}, \citenamefont {Ji}, \citenamefont {Yi}, \citenamefont {Li}, \citenamefont {Liu}, \citenamefont {Yu}, \citenamefont {Zhang} \emph {et~al.}}]{LNO327_SDW_R}%
  \BibitemOpen
  \bibfield  {author} {\bibinfo {author} {\bibfnamefont {Z.}~\bibnamefont {Liu}}, \bibinfo {author} {\bibfnamefont {H.}~\bibnamefont {Sun}}, \bibinfo {author} {\bibfnamefont {M.}~\bibnamefont {Huo}}, \bibinfo {author} {\bibfnamefont {X.}~\bibnamefont {Ma}}, \bibinfo {author} {\bibfnamefont {Y.}~\bibnamefont {Ji}}, \bibinfo {author} {\bibfnamefont {E.}~\bibnamefont {Yi}}, \bibinfo {author} {\bibfnamefont {L.}~\bibnamefont {Li}}, \bibinfo {author} {\bibfnamefont {H.}~\bibnamefont {Liu}}, \bibinfo {author} {\bibfnamefont {J.}~\bibnamefont {Yu}}, \bibinfo {author} {\bibfnamefont {Z.}~\bibnamefont {Zhang}}, \emph {et~al.},\ }\bibfield  {title} {\bibinfo {title} {{Evidence for charge and spin density waves in single crystals of La$_3$Ni$_2$O$_7$ and La$_3$Ni$_2$O$_6$}},\ }\href@noop {} {\bibfield  {journal} {\bibinfo  {journal} {Science China Physics, Mechanics \& Astronomy}\ }\textbf {\bibinfo {volume} {66}},\ \bibinfo {pages} {217411} (\bibinfo {year} {2023}{\natexlab{a}})}\BibitemShut {NoStop}%
\bibitem [{\citenamefont {TenHuisen}\ \emph {et~al.}(2025)\citenamefont {TenHuisen}, \citenamefont {Pan}, \citenamefont {Song}, \citenamefont {Baykusheva}, \citenamefont {Ferenc~Segedin}, \citenamefont {Goodge}, \citenamefont {Paik}, \citenamefont {Pelliciari}, \citenamefont {Bisogni}, \citenamefont {Gu}, \citenamefont {Agrestini}, \citenamefont {Nag}, \citenamefont {GarcÃ­a-FernÃ¡ndez}, \citenamefont {Zhou}, \citenamefont {Kourkoutis}, \citenamefont {Brooks}, \citenamefont {Mundy}, \citenamefont {Dean},\ and\ \citenamefont {Mitrano}}]{LNO4310_RIXS_PRB}%
  \BibitemOpen
  \bibfield  {author} {\bibinfo {author} {\bibfnamefont {S.~F.~R.}\ \bibnamefont {TenHuisen}}, \bibinfo {author} {\bibfnamefont {G.~A.}\ \bibnamefont {Pan}}, \bibinfo {author} {\bibfnamefont {Q.}~\bibnamefont {Song}}, \bibinfo {author} {\bibfnamefont {D.~R.}\ \bibnamefont {Baykusheva}}, \bibinfo {author} {\bibfnamefont {D.}~\bibnamefont {Ferenc~Segedin}}, \bibinfo {author} {\bibfnamefont {B.~H.}\ \bibnamefont {Goodge}}, \bibinfo {author} {\bibfnamefont {H.}~\bibnamefont {Paik}}, \bibinfo {author} {\bibfnamefont {J.}~\bibnamefont {Pelliciari}}, \bibinfo {author} {\bibfnamefont {V.}~\bibnamefont {Bisogni}}, \bibinfo {author} {\bibfnamefont {Y.}~\bibnamefont {Gu}}, \bibinfo {author} {\bibfnamefont {S.}~\bibnamefont {Agrestini}}, \bibinfo {author} {\bibfnamefont {A.}~\bibnamefont {Nag}}, \bibinfo {author} {\bibfnamefont {M.}~\bibnamefont {GarcÃ­a-FernÃ¡ndez}}, \bibinfo {author} {\bibfnamefont {K.-J.}\ \bibnamefont {Zhou}}, \bibinfo {author} {\bibfnamefont {L.~F.}\ \bibnamefont {Kourkoutis}}, \bibinfo {author}
  {\bibfnamefont {C.~M.}\ \bibnamefont {Brooks}}, \bibinfo {author} {\bibfnamefont {J.~A.}\ \bibnamefont {Mundy}}, \bibinfo {author} {\bibfnamefont {M.~P.~M.}\ \bibnamefont {Dean}},\ and\ \bibinfo {author} {\bibfnamefont {M.}~\bibnamefont {Mitrano}},\ }\bibfield  {title} {\bibinfo {title} {{Magnetic excitations in Ndn+1NinO3n+1 Ruddlesden-Popper nickelates observed via resonant inelastic x-ray scattering}},\ }\bibfield  {journal} {\bibinfo  {journal} {Physical Review B}\ }\textbf {\bibinfo {volume} {111}},\ \href {https://doi.org/10.1103/PhysRevB.111.165145} {10.1103/PhysRevB.111.165145} (\bibinfo {year} {2025})\BibitemShut {NoStop}%
\bibitem [{\citenamefont {Liu}\ \emph {et~al.}(2023{\natexlab{b}})\citenamefont {Liu}, \citenamefont {Mei}, \citenamefont {Ye}, \citenamefont {Chen},\ and\ \citenamefont {Yang}}]{LNO327_Theory_pair_PRL2023}%
  \BibitemOpen
  \bibfield  {author} {\bibinfo {author} {\bibfnamefont {Y.-B.}\ \bibnamefont {Liu}}, \bibinfo {author} {\bibfnamefont {J.-W.}\ \bibnamefont {Mei}}, \bibinfo {author} {\bibfnamefont {F.}~\bibnamefont {Ye}}, \bibinfo {author} {\bibfnamefont {W.-Q.}\ \bibnamefont {Chen}},\ and\ \bibinfo {author} {\bibfnamefont {F.}~\bibnamefont {Yang}},\ }\bibfield  {title} {\bibinfo {title} {{${s}^{\ifmmode\pm\else\textpm\fi{}}$-Wave Pairing and the Destructive Role of Apical-Oxygen Deficiencies in ${\mathrm{La}}_{3}{\mathrm{Ni}}_{2}{\mathrm{O}}_{7}$ under Pressure}},\ }\href {https://doi.org/10.1103/PhysRevLett.131.236002} {\bibfield  {journal} {\bibinfo  {journal} {Phys. Rev. Lett.}\ }\textbf {\bibinfo {volume} {131}},\ \bibinfo {pages} {236002} (\bibinfo {year} {2023}{\natexlab{b}})}\BibitemShut {NoStop}%
\bibitem [{\citenamefont {Zhang}\ \emph {et~al.}(2024{\natexlab{a}})\citenamefont {Zhang}, \citenamefont {Lin}, \citenamefont {Moreo}, \citenamefont {Maier},\ and\ \citenamefont {Dagotto}}]{LNO327_Theory_pair_SDW_NC2024}%
  \BibitemOpen
  \bibfield  {author} {\bibinfo {author} {\bibfnamefont {Y.}~\bibnamefont {Zhang}}, \bibinfo {author} {\bibfnamefont {L.~F.}\ \bibnamefont {Lin}}, \bibinfo {author} {\bibfnamefont {A.}~\bibnamefont {Moreo}}, \bibinfo {author} {\bibfnamefont {T.~A.}\ \bibnamefont {Maier}},\ and\ \bibinfo {author} {\bibfnamefont {E.}~\bibnamefont {Dagotto}},\ }\bibfield  {title} {\bibinfo {title} {{Structural phase transition, s(+/-)-wave pairing, and magnetic stripe order in bilayered superconductor La(3)Ni(2)O(7) under pressure}},\ }\href {https://doi.org/10.1038/s41467-024-46622-z} {\bibfield  {journal} {\bibinfo  {journal} {Nat Commun}\ }\textbf {\bibinfo {volume} {15}},\ \bibinfo {pages} {2470} (\bibinfo {year} {2024}{\natexlab{a}})}\BibitemShut {NoStop}%
\bibitem [{\citenamefont {Zhang}\ \emph {et~al.}(2023)\citenamefont {Zhang}, \citenamefont {Lin}, \citenamefont {Moreo},\ and\ \citenamefont {Dagotto}}]{LNO327_Theory_press_PRB2023}%
  \BibitemOpen
  \bibfield  {author} {\bibinfo {author} {\bibfnamefont {Y.}~\bibnamefont {Zhang}}, \bibinfo {author} {\bibfnamefont {L.-F.}\ \bibnamefont {Lin}}, \bibinfo {author} {\bibfnamefont {A.}~\bibnamefont {Moreo}},\ and\ \bibinfo {author} {\bibfnamefont {E.}~\bibnamefont {Dagotto}},\ }\bibfield  {title} {\bibinfo {title} {{Electronic structure, dimer physics, orbital-selective behavior, and magnetic tendencies in the bilayer nickelate superconductor $La_{3}Ni_{2}O_{7}$ under pressure}},\ }\href {https://doi.org/10.1103/PhysRevB.108.L180510} {\bibfield  {journal} {\bibinfo  {journal} {Phys. Rev. B}\ }\textbf {\bibinfo {volume} {108}},\ \bibinfo {pages} {L180510} (\bibinfo {year} {2023})}\BibitemShut {NoStop}%
\bibitem [{\citenamefont {Christiansson}\ \emph {et~al.}(2023)\citenamefont {Christiansson}, \citenamefont {Petocchi},\ and\ \citenamefont {Werner}}]{LNO327_Theory_ele_struct_PRL2023}%
  \BibitemOpen
  \bibfield  {author} {\bibinfo {author} {\bibfnamefont {V.}~\bibnamefont {Christiansson}}, \bibinfo {author} {\bibfnamefont {F.}~\bibnamefont {Petocchi}},\ and\ \bibinfo {author} {\bibfnamefont {P.}~\bibnamefont {Werner}},\ }\bibfield  {title} {\bibinfo {title} {{Correlated Electronic Structure of La$_3$Ni$_2$O$_7$ under Pressure}},\ }\bibfield  {journal} {\bibinfo  {journal} {Physical Review Letters}\ }\textbf {\bibinfo {volume} {131}},\ \href {https://doi.org/10.1103/PhysRevLett.131.206501} {10.1103/PhysRevLett.131.206501} (\bibinfo {year} {2023})\BibitemShut {NoStop}%
\bibitem [{\citenamefont {Luo}\ \emph {et~al.}(2024)\citenamefont {Luo}, \citenamefont {Lv}, \citenamefont {Wang}, \citenamefont {WÃº},\ and\ \citenamefont {Yao}}]{LNO327_theory_YDX_npj2024}%
  \BibitemOpen
  \bibfield  {author} {\bibinfo {author} {\bibfnamefont {Z.}~\bibnamefont {Luo}}, \bibinfo {author} {\bibfnamefont {B.}~\bibnamefont {Lv}}, \bibinfo {author} {\bibfnamefont {M.}~\bibnamefont {Wang}}, \bibinfo {author} {\bibfnamefont {W.}~\bibnamefont {WÃº}},\ and\ \bibinfo {author} {\bibfnamefont {D.-X.}\ \bibnamefont {Yao}},\ }\bibfield  {title} {\bibinfo {title} {{High-TC superconductivity in La3Ni2O7 based on the bilayer two-orbital t-J model}},\ }\href {https://doi.org/10.1038/s41535-024-00668-w} {\bibfield  {journal} {\bibinfo  {journal} {npj Quantum Materials}\ }\textbf {\bibinfo {volume} {9}},\ \bibinfo {pages} {61} (\bibinfo {year} {2024})}\BibitemShut {NoStop}%
\bibitem [{\citenamefont {Zhang}\ \emph {et~al.}(2024{\natexlab{b}})\citenamefont {Zhang}, \citenamefont {Sun}, \citenamefont {Liu}, \citenamefont {Liu}, \citenamefont {Chen},\ and\ \citenamefont {Yang}}]{LNO4310_theory_SC_PRB2024}%
  \BibitemOpen
  \bibfield  {author} {\bibinfo {author} {\bibfnamefont {M.}~\bibnamefont {Zhang}}, \bibinfo {author} {\bibfnamefont {H.}~\bibnamefont {Sun}}, \bibinfo {author} {\bibfnamefont {Y.-B.}\ \bibnamefont {Liu}}, \bibinfo {author} {\bibfnamefont {Q.}~\bibnamefont {Liu}}, \bibinfo {author} {\bibfnamefont {W.-Q.}\ \bibnamefont {Chen}},\ and\ \bibinfo {author} {\bibfnamefont {F.}~\bibnamefont {Yang}},\ }\bibfield  {title} {\bibinfo {title} {{sÂ±wave superconductivity in pressurized La$_4$Ni$_3$O$_{10}$}},\ }\bibfield  {journal} {\bibinfo  {journal} {Physical Review B}\ }\textbf {\bibinfo {volume} {110}},\ \href {https://doi.org/10.1103/PhysRevB.110.L180501} {10.1103/PhysRevB.110.L180501} (\bibinfo {year} {2024}{\natexlab{b}})\BibitemShut {NoStop}%
\bibitem [{\citenamefont {Lu}\ \emph {et~al.}(2025)\citenamefont {Lu}, \citenamefont {Pan}, \citenamefont {Yang},\ and\ \citenamefont {Wu}}]{LNO4310_theory_SC_PRB2025}%
  \BibitemOpen
  \bibfield  {author} {\bibinfo {author} {\bibfnamefont {C.}~\bibnamefont {Lu}}, \bibinfo {author} {\bibfnamefont {Z.}~\bibnamefont {Pan}}, \bibinfo {author} {\bibfnamefont {F.}~\bibnamefont {Yang}},\ and\ \bibinfo {author} {\bibfnamefont {C.}~\bibnamefont {Wu}},\ }\bibfield  {title} {\bibinfo {title} {{Superconductivity in La$_4$Ni$_3$O$_{10}$ under pressure}},\ }\bibfield  {journal} {\bibinfo  {journal} {Physical Review B}\ }\textbf {\bibinfo {volume} {111}},\ \href {https://doi.org/10.1103/PhysRevB.111.134515} {10.1103/PhysRevB.111.134515} (\bibinfo {year} {2025})\BibitemShut {NoStop}%
\bibitem [{\citenamefont {Huang}\ and\ \citenamefont {Zhou}(2024)}]{LNO4310_theory_interlay_PRB2024}%
  \BibitemOpen
  \bibfield  {author} {\bibinfo {author} {\bibfnamefont {J.}~\bibnamefont {Huang}}\ and\ \bibinfo {author} {\bibfnamefont {T.}~\bibnamefont {Zhou}},\ }\bibfield  {title} {\bibinfo {title} {{Interlayer pairing-induced partially gapped Fermi surface in trilayer La$_4$Ni$_3$O$_{10}$ superconductors}},\ }\bibfield  {journal} {\bibinfo  {journal} {Physical Review B}\ }\textbf {\bibinfo {volume} {110}},\ \href {https://doi.org/10.1103/PhysRevB.110.L060506} {10.1103/PhysRevB.110.L060506} (\bibinfo {year} {2024})\BibitemShut {NoStop}%
\bibitem [{\citenamefont {Ryee}\ \emph {et~al.}(2024)\citenamefont {Ryee}, \citenamefont {Witt},\ and\ \citenamefont {Wehling}}]{LNO327_theory_dxy_pair_PRL2024}%
  \BibitemOpen
  \bibfield  {author} {\bibinfo {author} {\bibfnamefont {S.}~\bibnamefont {Ryee}}, \bibinfo {author} {\bibfnamefont {N.}~\bibnamefont {Witt}},\ and\ \bibinfo {author} {\bibfnamefont {T.~O.}\ \bibnamefont {Wehling}},\ }\bibfield  {title} {\bibinfo {title} {{Quenched Pair Breaking by Interlayer Correlations as a Key to Superconductivity in ${\mathrm{La}}_{3}{\mathrm{Ni}}_{2}{\mathrm{O}}_{7}$}},\ }\href {https://doi.org/10.1103/PhysRevLett.133.096002} {\bibfield  {journal} {\bibinfo  {journal} {Phys. Rev. Lett.}\ }\textbf {\bibinfo {volume} {133}},\ \bibinfo {pages} {096002} (\bibinfo {year} {2024})}\BibitemShut {NoStop}%
\bibitem [{\citenamefont {Qu}\ \emph {et~al.}(2024)\citenamefont {Qu}, \citenamefont {Qu}, \citenamefont {Chen}, \citenamefont {Wu}, \citenamefont {Yang}, \citenamefont {Li},\ and\ \citenamefont {Su}}]{LNO327_theory_dxy_pair_2_PRL2024}%
  \BibitemOpen
  \bibfield  {author} {\bibinfo {author} {\bibfnamefont {X.-Z.}\ \bibnamefont {Qu}}, \bibinfo {author} {\bibfnamefont {D.-W.}\ \bibnamefont {Qu}}, \bibinfo {author} {\bibfnamefont {J.}~\bibnamefont {Chen}}, \bibinfo {author} {\bibfnamefont {C.}~\bibnamefont {Wu}}, \bibinfo {author} {\bibfnamefont {F.}~\bibnamefont {Yang}}, \bibinfo {author} {\bibfnamefont {W.}~\bibnamefont {Li}},\ and\ \bibinfo {author} {\bibfnamefont {G.}~\bibnamefont {Su}},\ }\bibfield  {title} {\bibinfo {title} {{Bilayer ${t\text{\ensuremath{-}}J\text{\ensuremath{-}}J}_{\ensuremath{\perp}}$ Model and Magnetically Mediated Pairing in the Pressurized Nickelate ${\mathrm{La}}_{3}{\mathrm{Ni}}_{2}{\mathrm{O}}_{7}$}},\ }\href {https://doi.org/10.1103/PhysRevLett.132.036502} {\bibfield  {journal} {\bibinfo  {journal} {Phys. Rev. Lett.}\ }\textbf {\bibinfo {volume} {132}},\ \bibinfo {pages} {036502} (\bibinfo {year} {2024})}\BibitemShut {NoStop}%
\bibitem [{\citenamefont {Chen}\ \emph {et~al.}(2024{\natexlab{b}})\citenamefont {Chen}, \citenamefont {Yang},\ and\ \citenamefont {Li}}]{LNO327_Theory_dxy_pair_PRB2024}%
  \BibitemOpen
  \bibfield  {author} {\bibinfo {author} {\bibfnamefont {J.}~\bibnamefont {Chen}}, \bibinfo {author} {\bibfnamefont {F.}~\bibnamefont {Yang}},\ and\ \bibinfo {author} {\bibfnamefont {W.}~\bibnamefont {Li}},\ }\bibfield  {title} {\bibinfo {title} {{Orbital-selective superconductivity in the pressurized bilayer nickelate ${\mathrm{La}}_{3}{\mathrm{Ni}}_{2}{\mathrm{O}}_{7}$: An infinite projected entangled-pair state study}},\ }\href {https://doi.org/10.1103/PhysRevB.110.L041111} {\bibfield  {journal} {\bibinfo  {journal} {Phys. Rev. B}\ }\textbf {\bibinfo {volume} {110}},\ \bibinfo {pages} {L041111} (\bibinfo {year} {2024}{\natexlab{b}})}\BibitemShut {NoStop}%
\bibitem [{\citenamefont {Yang}\ \emph {et~al.}(2024{\natexlab{b}})\citenamefont {Yang}, \citenamefont {Jiang}, \citenamefont {Wang}, \citenamefont {Lu},\ and\ \citenamefont {Wang}}]{LNO4310_Theory_dz_pair_PRB2024}%
  \BibitemOpen
  \bibfield  {author} {\bibinfo {author} {\bibfnamefont {Q.-G.}\ \bibnamefont {Yang}}, \bibinfo {author} {\bibfnamefont {K.-Y.}\ \bibnamefont {Jiang}}, \bibinfo {author} {\bibfnamefont {D.}~\bibnamefont {Wang}}, \bibinfo {author} {\bibfnamefont {H.-Y.}\ \bibnamefont {Lu}},\ and\ \bibinfo {author} {\bibfnamefont {Q.-H.}\ \bibnamefont {Wang}},\ }\bibfield  {title} {\bibinfo {title} {{Effective model and ${s}_{\ifmmode\pm\else\textpm\fi{}}$-wave superconductivity in trilayer nickelate ${\mathrm{La}}_{4}{\mathrm{Ni}}_{3}{\mathrm{O}}_{10}$}},\ }\href {https://doi.org/10.1103/PhysRevB.109.L220506} {\bibfield  {journal} {\bibinfo  {journal} {Phys. Rev. B}\ }\textbf {\bibinfo {volume} {109}},\ \bibinfo {pages} {L220506} (\bibinfo {year} {2024}{\natexlab{b}})}\BibitemShut {NoStop}%
\bibitem [{\citenamefont {Qin}\ and\ \citenamefont {Yang}(2023)}]{LNO327_theory_spinsiglet_PRB2023}%
  \BibitemOpen
  \bibfield  {author} {\bibinfo {author} {\bibfnamefont {Q.}~\bibnamefont {Qin}}\ and\ \bibinfo {author} {\bibfnamefont {Y.-f.}\ \bibnamefont {Yang}},\ }\bibfield  {title} {\bibinfo {title} {{High-${T}_{c}$ superconductivity by mobilizing local spin singlets and possible route to higher ${T}_{c}$ in pressurized ${\mathrm{La}}_{3}{\mathrm{Ni}}_{2}{\mathrm{O}}_{7}$}},\ }\href {https://doi.org/10.1103/PhysRevB.108.L140504} {\bibfield  {journal} {\bibinfo  {journal} {Phys. Rev. B}\ }\textbf {\bibinfo {volume} {108}},\ \bibinfo {pages} {L140504} (\bibinfo {year} {2023})}\BibitemShut {NoStop}%
\bibitem [{\citenamefont {Dong}\ \emph {et~al.}(2024)\citenamefont {Dong}, \citenamefont {Huo}, \citenamefont {Li}, \citenamefont {Li}, \citenamefont {Li}, \citenamefont {Sun}, \citenamefont {Gu}, \citenamefont {Lu}, \citenamefont {Wang}, \citenamefont {Wang},\ and\ \citenamefont {Chen}}]{EELS_327_Nat}%
  \BibitemOpen
  \bibfield  {author} {\bibinfo {author} {\bibfnamefont {Z.}~\bibnamefont {Dong}}, \bibinfo {author} {\bibfnamefont {M.}~\bibnamefont {Huo}}, \bibinfo {author} {\bibfnamefont {J.}~\bibnamefont {Li}}, \bibinfo {author} {\bibfnamefont {J.}~\bibnamefont {Li}}, \bibinfo {author} {\bibfnamefont {P.}~\bibnamefont {Li}}, \bibinfo {author} {\bibfnamefont {H.}~\bibnamefont {Sun}}, \bibinfo {author} {\bibfnamefont {L.}~\bibnamefont {Gu}}, \bibinfo {author} {\bibfnamefont {Y.}~\bibnamefont {Lu}}, \bibinfo {author} {\bibfnamefont {M.}~\bibnamefont {Wang}}, \bibinfo {author} {\bibfnamefont {Y.}~\bibnamefont {Wang}},\ and\ \bibinfo {author} {\bibfnamefont {Z.}~\bibnamefont {Chen}},\ }\bibfield  {title} {\bibinfo {title} {{Visualization of oxygen vacancies and self-doped ligand holes in La$_3$Ni$_2$O$_{7-\delta}$}},\ }\href {https://doi.org/10.1038/s41586-024-07482-1} {\bibfield  {journal} {\bibinfo  {journal} {Nature}\ }\textbf {\bibinfo {volume} {630}},\ \bibinfo {pages} {847} (\bibinfo {year} {2024})}\BibitemShut
  {NoStop}%
\bibitem [{\citenamefont {Dong}\ \emph {et~al.}(2025)\citenamefont {Dong}, \citenamefont {Wang}, \citenamefont {Wang}, \citenamefont {Dong}, \citenamefont {Gu}, \citenamefont {Xu}, \citenamefont {Cheng}, \citenamefont {Chen},\ and\ \citenamefont {Wang}}]{LPNO_EELS_Dong_NM}%
  \BibitemOpen
  \bibfield  {author} {\bibinfo {author} {\bibfnamefont {Z.}~\bibnamefont {Dong}}, \bibinfo {author} {\bibfnamefont {G.}~\bibnamefont {Wang}}, \bibinfo {author} {\bibfnamefont {N.}~\bibnamefont {Wang}}, \bibinfo {author} {\bibfnamefont {W.-H.}\ \bibnamefont {Dong}}, \bibinfo {author} {\bibfnamefont {L.}~\bibnamefont {Gu}}, \bibinfo {author} {\bibfnamefont {Y.}~\bibnamefont {Xu}}, \bibinfo {author} {\bibfnamefont {J.}~\bibnamefont {Cheng}}, \bibinfo {author} {\bibfnamefont {Z.}~\bibnamefont {Chen}},\ and\ \bibinfo {author} {\bibfnamefont {Y.}~\bibnamefont {Wang}},\ }\bibfield  {title} {\bibinfo {title} {{Interstitial oxygen order and its competition with superconductivity in La$_2$PrNi$_2$O$_{7+\delta}$}},\ }\bibfield  {journal} {\bibinfo  {journal} {arXiv preprint arXiv:2508.03414}\ }\href {https://doi.org/doi.org/10.48550/arXiv.2508.03414} {doi.org/10.48550/arXiv.2508.03414} (\bibinfo {year} {2025})\BibitemShut {NoStop}%
\bibitem [{\citenamefont {Saini}\ \emph {et~al.}(1996)\citenamefont {Saini}, \citenamefont {Venkatesh}, \citenamefont {Srivastava}, \citenamefont {Sekhar}, \citenamefont {Garg}, \citenamefont {Tjeng}, \citenamefont {Chen}, \citenamefont {Menovsky},\ and\ \citenamefont {Franse}}]{XAS_pol_1996}%
  \BibitemOpen
  \bibfield  {author} {\bibinfo {author} {\bibfnamefont {N.~L.}\ \bibnamefont {Saini}}, \bibinfo {author} {\bibfnamefont {S.}~\bibnamefont {Venkatesh}}, \bibinfo {author} {\bibfnamefont {P.}~\bibnamefont {Srivastava}}, \bibinfo {author} {\bibfnamefont {B.~R.}\ \bibnamefont {Sekhar}}, \bibinfo {author} {\bibfnamefont {K.~B.}\ \bibnamefont {Garg}}, \bibinfo {author} {\bibfnamefont {L.~H.}\ \bibnamefont {Tjeng}}, \bibinfo {author} {\bibfnamefont {C.~T.}\ \bibnamefont {Chen}}, \bibinfo {author} {\bibfnamefont {A.}~\bibnamefont {Menovsky}},\ and\ \bibinfo {author} {\bibfnamefont {J.~J.~M.}\ \bibnamefont {Franse}},\ }\bibfield  {title} {\bibinfo {title} {{Polarized x-ray absorption spectroscopy study of the symmetry of unoccupied electronic states near the Fermi level in the system}},\ }\href {https://doi.org/10.1088/0953-8984/8/14/020} {\bibfield  {journal} {\bibinfo  {journal} {Journal of Physics: Condensed Matter}\ }\textbf {\bibinfo {volume} {8}},\ \bibinfo {pages} {2467} (\bibinfo {year} {1996})}\BibitemShut
  {NoStop}%
\bibitem [{\citenamefont {Ament}\ \emph {et~al.}(2011)\citenamefont {Ament}, \citenamefont {van Veenendaal}, \citenamefont {Devereaux}, \citenamefont {Hill},\ and\ \citenamefont {van~den Brink}}]{RIXS_REXS_tech_1}%
  \BibitemOpen
  \bibfield  {author} {\bibinfo {author} {\bibfnamefont {L.~J.~P.}\ \bibnamefont {Ament}}, \bibinfo {author} {\bibfnamefont {M.}~\bibnamefont {van Veenendaal}}, \bibinfo {author} {\bibfnamefont {T.~P.}\ \bibnamefont {Devereaux}}, \bibinfo {author} {\bibfnamefont {J.~P.}\ \bibnamefont {Hill}},\ and\ \bibinfo {author} {\bibfnamefont {J.}~\bibnamefont {van~den Brink}},\ }\bibfield  {title} {\bibinfo {title} {{Resonant inelastic x-ray scattering studies of elementary excitations}},\ }\href {https://doi.org/10.1103/RevModPhys.83.705} {\bibfield  {journal} {\bibinfo  {journal} {Reviews of Modern Physics}\ }\textbf {\bibinfo {volume} {83}},\ \bibinfo {pages} {705} (\bibinfo {year} {2011})}\BibitemShut {NoStop}%
\bibitem [{\citenamefont {Li}\ \emph {et~al.}(2024{\natexlab{b}})\citenamefont {Li}, \citenamefont {Chen}, \citenamefont {Huang}, \citenamefont {Han}, \citenamefont {Huo}, \citenamefont {Huang}, \citenamefont {Ma}, \citenamefont {Qiu}, \citenamefont {Chen}, \citenamefont {Hu}, \citenamefont {Chen}, \citenamefont {Xie}, \citenamefont {Shen}, \citenamefont {Sun}, \citenamefont {Yao},\ and\ \citenamefont {Wang}}]{LNO4310_struct_pre_WM}%
  \BibitemOpen
  \bibfield  {author} {\bibinfo {author} {\bibfnamefont {J.}~\bibnamefont {Li}}, \bibinfo {author} {\bibfnamefont {C.-Q.}\ \bibnamefont {Chen}}, \bibinfo {author} {\bibfnamefont {C.}~\bibnamefont {Huang}}, \bibinfo {author} {\bibfnamefont {Y.}~\bibnamefont {Han}}, \bibinfo {author} {\bibfnamefont {M.}~\bibnamefont {Huo}}, \bibinfo {author} {\bibfnamefont {X.}~\bibnamefont {Huang}}, \bibinfo {author} {\bibfnamefont {P.}~\bibnamefont {Ma}}, \bibinfo {author} {\bibfnamefont {Z.}~\bibnamefont {Qiu}}, \bibinfo {author} {\bibfnamefont {J.}~\bibnamefont {Chen}}, \bibinfo {author} {\bibfnamefont {X.}~\bibnamefont {Hu}}, \bibinfo {author} {\bibfnamefont {L.}~\bibnamefont {Chen}}, \bibinfo {author} {\bibfnamefont {T.}~\bibnamefont {Xie}}, \bibinfo {author} {\bibfnamefont {B.}~\bibnamefont {Shen}}, \bibinfo {author} {\bibfnamefont {H.}~\bibnamefont {Sun}}, \bibinfo {author} {\bibfnamefont {D.-X.}\ \bibnamefont {Yao}},\ and\ \bibinfo {author} {\bibfnamefont {M.}~\bibnamefont {Wang}},\ }\bibfield  {title} {\bibinfo
  {title} {{Structural transition, electric transport, and electronic structures in the compressed trilayer nickelate La$_4$Ni$_3$O$_{10}$}},\ }\bibfield  {journal} {\bibinfo  {journal} {Science China Physics, Mechanics\& Astronomy}\ }\textbf {\bibinfo {volume} {67}},\ \href {https://doi.org/10.1007/s11433-023-2329-x} {10.1007/s11433-023-2329-x} (\bibinfo {year} {2024}{\natexlab{b}})\BibitemShut {NoStop}%
\bibitem [{\citenamefont {Hepting}\ \emph {et~al.}(2020)\citenamefont {Hepting}, \citenamefont {Li}, \citenamefont {Jia}, \citenamefont {Lu}, \citenamefont {Paris}, \citenamefont {Tseng}, \citenamefont {Feng}, \citenamefont {Osada}, \citenamefont {Been}, \citenamefont {Hikita}, \citenamefont {Chuang}, \citenamefont {Hussain}, \citenamefont {Zhou}, \citenamefont {Nag}, \citenamefont {Garcia-Fernandez}, \citenamefont {Rossi}, \citenamefont {Huang}, \citenamefont {Huang}, \citenamefont {Shen}, \citenamefont {Schmitt}, \citenamefont {Hwang}, \citenamefont {Moritz}, \citenamefont {Zaanen}, \citenamefont {Devereaux},\ and\ \citenamefont {Lee}}]{LNO112_band_XASXES_NM2020}%
  \BibitemOpen
  \bibfield  {author} {\bibinfo {author} {\bibfnamefont {M.}~\bibnamefont {Hepting}}, \bibinfo {author} {\bibfnamefont {D.}~\bibnamefont {Li}}, \bibinfo {author} {\bibfnamefont {C.~J.}\ \bibnamefont {Jia}}, \bibinfo {author} {\bibfnamefont {H.}~\bibnamefont {Lu}}, \bibinfo {author} {\bibfnamefont {E.}~\bibnamefont {Paris}}, \bibinfo {author} {\bibfnamefont {Y.}~\bibnamefont {Tseng}}, \bibinfo {author} {\bibfnamefont {X.}~\bibnamefont {Feng}}, \bibinfo {author} {\bibfnamefont {M.}~\bibnamefont {Osada}}, \bibinfo {author} {\bibfnamefont {E.}~\bibnamefont {Been}}, \bibinfo {author} {\bibfnamefont {Y.}~\bibnamefont {Hikita}}, \bibinfo {author} {\bibfnamefont {Y.~D.}\ \bibnamefont {Chuang}}, \bibinfo {author} {\bibfnamefont {Z.}~\bibnamefont {Hussain}}, \bibinfo {author} {\bibfnamefont {K.~J.}\ \bibnamefont {Zhou}}, \bibinfo {author} {\bibfnamefont {A.}~\bibnamefont {Nag}}, \bibinfo {author} {\bibfnamefont {M.}~\bibnamefont {Garcia-Fernandez}}, \bibinfo {author} {\bibfnamefont {M.}~\bibnamefont {Rossi}}, \bibinfo
  {author} {\bibfnamefont {H.~Y.}\ \bibnamefont {Huang}}, \bibinfo {author} {\bibfnamefont {D.~J.}\ \bibnamefont {Huang}}, \bibinfo {author} {\bibfnamefont {Z.~X.}\ \bibnamefont {Shen}}, \bibinfo {author} {\bibfnamefont {T.}~\bibnamefont {Schmitt}}, \bibinfo {author} {\bibfnamefont {H.~Y.}\ \bibnamefont {Hwang}}, \bibinfo {author} {\bibfnamefont {B.}~\bibnamefont {Moritz}}, \bibinfo {author} {\bibfnamefont {J.}~\bibnamefont {Zaanen}}, \bibinfo {author} {\bibfnamefont {T.~P.}\ \bibnamefont {Devereaux}},\ and\ \bibinfo {author} {\bibfnamefont {W.~S.}\ \bibnamefont {Lee}},\ }\bibfield  {title} {\bibinfo {title} {{Electronic structure of the parent compound of superconducting infinite-layer nickelates}},\ }\href {https://doi.org/10.1038/s41563-019-0585-z} {\bibfield  {journal} {\bibinfo  {journal} {Nat Mater}\ }\textbf {\bibinfo {volume} {19}},\ \bibinfo {pages} {381} (\bibinfo {year} {2020})}\BibitemShut {NoStop}%
\bibitem [{\citenamefont {Timrov}\ \emph {et~al.}(2020)\citenamefont {Timrov}, \citenamefont {Agrawal}, \citenamefont {Zhang}, \citenamefont {Erat}, \citenamefont {Liu}, \citenamefont {Braun}, \citenamefont {Cococcioni}, \citenamefont {Calandra}, \citenamefont {Marzari},\ and\ \citenamefont {Passerone}}]{DFT_XAS_mismath_1}%
  \BibitemOpen
  \bibfield  {author} {\bibinfo {author} {\bibfnamefont {I.}~\bibnamefont {Timrov}}, \bibinfo {author} {\bibfnamefont {P.}~\bibnamefont {Agrawal}}, \bibinfo {author} {\bibfnamefont {X.}~\bibnamefont {Zhang}}, \bibinfo {author} {\bibfnamefont {S.}~\bibnamefont {Erat}}, \bibinfo {author} {\bibfnamefont {R.}~\bibnamefont {Liu}}, \bibinfo {author} {\bibfnamefont {A.}~\bibnamefont {Braun}}, \bibinfo {author} {\bibfnamefont {M.}~\bibnamefont {Cococcioni}}, \bibinfo {author} {\bibfnamefont {M.}~\bibnamefont {Calandra}}, \bibinfo {author} {\bibfnamefont {N.}~\bibnamefont {Marzari}},\ and\ \bibinfo {author} {\bibfnamefont {D.}~\bibnamefont {Passerone}},\ }\bibfield  {title} {\bibinfo {title} {{Electronic structure of pristine and Ni-substituted $\mathrm{La}\mathrm{Fe}{\mathrm{O}}_{3}$ from near edge x-ray absorption fine structure experiments and first-principles simulations}},\ }\href {https://doi.org/10.1103/PhysRevResearch.2.033265} {\bibfield  {journal} {\bibinfo  {journal} {Phys. Rev. Res.}\ }\textbf {\bibinfo
  {volume} {2}},\ \bibinfo {pages} {033265} (\bibinfo {year} {2020})}\BibitemShut {NoStop}%
\bibitem [{\citenamefont {Vinson}\ \emph {et~al.}(2011)\citenamefont {Vinson}, \citenamefont {Rehr}, \citenamefont {Kas},\ and\ \citenamefont {Shirley}}]{DFT_XAS_mismath_2}%
  \BibitemOpen
  \bibfield  {author} {\bibinfo {author} {\bibfnamefont {J.}~\bibnamefont {Vinson}}, \bibinfo {author} {\bibfnamefont {J.~J.}\ \bibnamefont {Rehr}}, \bibinfo {author} {\bibfnamefont {J.~J.}\ \bibnamefont {Kas}},\ and\ \bibinfo {author} {\bibfnamefont {E.~L.}\ \bibnamefont {Shirley}},\ }\bibfield  {title} {\bibinfo {title} {{Bethe-Salpeter equation calculations of core excitation spectra}},\ }\href {https://doi.org/10.1103/PhysRevB.83.115106} {\bibfield  {journal} {\bibinfo  {journal} {Phys. Rev. B}\ }\textbf {\bibinfo {volume} {83}},\ \bibinfo {pages} {115106} (\bibinfo {year} {2011})}\BibitemShut {NoStop}%
\bibitem [{\citenamefont {Deswal}\ \emph {et~al.}(2024)\citenamefont {Deswal}, \citenamefont {Kumar}, \citenamefont {Rout}, \citenamefont {Singh},\ and\ \citenamefont {Kumar}}]{Raman_india}%
  \BibitemOpen
  \bibfield  {author} {\bibinfo {author} {\bibfnamefont {S.}~\bibnamefont {Deswal}}, \bibinfo {author} {\bibfnamefont {D.}~\bibnamefont {Kumar}}, \bibinfo {author} {\bibfnamefont {D.}~\bibnamefont {Rout}}, \bibinfo {author} {\bibfnamefont {S.}~\bibnamefont {Singh}},\ and\ \bibinfo {author} {\bibfnamefont {P.}~\bibnamefont {Kumar}},\ }\href {https://arxiv.org/abs/2411.13933} {\bibinfo {title} {{Dynamics of electron-electron correlated to electron-phonon coupled phase progression in trilayer nickelate La$_4$Ni$_3$O$_{10}$}}} (\bibinfo {year} {2024}),\ \Eprint {https://arxiv.org/abs/2411.13933} {arXiv:2411.13933 [cond-mat.str-el]} \BibitemShut {NoStop}%
\bibitem [{\citenamefont {Ament}\ \emph {et~al.}(2009)\citenamefont {Ament}, \citenamefont {Ghiringhelli}, \citenamefont {Sala}, \citenamefont {Braicovich},\ and\ \citenamefont {van~den Brink}}]{Magnon_RIXS_PRL}%
  \BibitemOpen
  \bibfield  {author} {\bibinfo {author} {\bibfnamefont {L.~J.~P.}\ \bibnamefont {Ament}}, \bibinfo {author} {\bibfnamefont {G.}~\bibnamefont {Ghiringhelli}}, \bibinfo {author} {\bibfnamefont {M.~M.}\ \bibnamefont {Sala}}, \bibinfo {author} {\bibfnamefont {L.}~\bibnamefont {Braicovich}},\ and\ \bibinfo {author} {\bibfnamefont {J.}~\bibnamefont {van~den Brink}},\ }\bibfield  {title} {\bibinfo {title} {{Theoretical Demonstration of How the Dispersion of Magnetic Excitations in Cuprate Compounds can be Determined Using Resonant Inelastic X-Ray Scattering}},\ }\bibfield  {journal} {\bibinfo  {journal} {Physical Review Letters}\ }\textbf {\bibinfo {volume} {103}},\ \href {https://doi.org/10.1103/PhysRevLett.103.117003} {10.1103/PhysRevLett.103.117003} (\bibinfo {year} {2009})\BibitemShut {NoStop}%
\bibitem [{\citenamefont {Bisogni}\ \emph {et~al.}(2012)\citenamefont {Bisogni}, \citenamefont {Simonelli}, \citenamefont {Ament}, \citenamefont {Forte}, \citenamefont {Moretti~Sala}, \citenamefont {Minola}, \citenamefont {Huotari}, \citenamefont {van~den Brink}, \citenamefont {Ghiringhelli}, \citenamefont {Brookes},\ and\ \citenamefont {Braicovich}}]{cuprate_OK_bimagnon}%
  \BibitemOpen
  \bibfield  {author} {\bibinfo {author} {\bibfnamefont {V.}~\bibnamefont {Bisogni}}, \bibinfo {author} {\bibfnamefont {L.}~\bibnamefont {Simonelli}}, \bibinfo {author} {\bibfnamefont {L.~J.~P.}\ \bibnamefont {Ament}}, \bibinfo {author} {\bibfnamefont {F.}~\bibnamefont {Forte}}, \bibinfo {author} {\bibfnamefont {M.}~\bibnamefont {Moretti~Sala}}, \bibinfo {author} {\bibfnamefont {M.}~\bibnamefont {Minola}}, \bibinfo {author} {\bibfnamefont {S.}~\bibnamefont {Huotari}}, \bibinfo {author} {\bibfnamefont {J.}~\bibnamefont {van~den Brink}}, \bibinfo {author} {\bibfnamefont {G.}~\bibnamefont {Ghiringhelli}}, \bibinfo {author} {\bibfnamefont {N.~B.}\ \bibnamefont {Brookes}},\ and\ \bibinfo {author} {\bibfnamefont {L.}~\bibnamefont {Braicovich}},\ }\bibfield  {title} {\bibinfo {title} {{Bimagnon studies in cuprates with resonant inelastic x-ray scattering at the OK edge. I. Assessment on $La_2CuO_4$ and comparison with the excitation at Cu L3 and Cu K edges}},\ }\bibfield  {journal} {\bibinfo  {journal} {Physical
  Review B}\ }\textbf {\bibinfo {volume} {85}},\ \href {https://doi.org/10.1103/PhysRevB.85.214527} {10.1103/PhysRevB.85.214527} (\bibinfo {year} {2012})\BibitemShut {NoStop}%
\bibitem [{\citenamefont {Singh}\ \emph {et~al.}(1989)\citenamefont {Singh}, \citenamefont {Fleury}, \citenamefont {Lyons},\ and\ \citenamefont {Sulewski}}]{Cuprate_bimagnon_Raman}%
  \BibitemOpen
  \bibfield  {author} {\bibinfo {author} {\bibfnamefont {R.~R.~P.}\ \bibnamefont {Singh}}, \bibinfo {author} {\bibfnamefont {P.~A.}\ \bibnamefont {Fleury}}, \bibinfo {author} {\bibfnamefont {K.~B.}\ \bibnamefont {Lyons}},\ and\ \bibinfo {author} {\bibfnamefont {P.~E.}\ \bibnamefont {Sulewski}},\ }\bibfield  {title} {\bibinfo {title} {{Quantitative Determination of Quantum Fluctuations in the Spin-1/2 Planar Antiferromagnet}},\ }\href {https://doi.org/10.1103/PhysRevLett.62.2736} {\bibfield  {journal} {\bibinfo  {journal} {Physical Review Letters}\ }\textbf {\bibinfo {volume} {62}},\ \bibinfo {pages} {2736} (\bibinfo {year} {1989})}\BibitemShut {NoStop}%
\bibitem [{\citenamefont {Chen}\ \emph {et~al.}(2024{\natexlab{c}})\citenamefont {Chen}, \citenamefont {Luo}, \citenamefont {Wang}, \citenamefont {W\'u},\ and\ \citenamefont {Yao}}]{LNO4310_theory_nobounding_YDX_PRB2024}%
  \BibitemOpen
  \bibfield  {author} {\bibinfo {author} {\bibfnamefont {C.-Q.}\ \bibnamefont {Chen}}, \bibinfo {author} {\bibfnamefont {Z.}~\bibnamefont {Luo}}, \bibinfo {author} {\bibfnamefont {M.}~\bibnamefont {Wang}}, \bibinfo {author} {\bibfnamefont {W.}~\bibnamefont {W\'u}},\ and\ \bibinfo {author} {\bibfnamefont {D.-X.}\ \bibnamefont {Yao}},\ }\bibfield  {title} {\bibinfo {title} {{Trilayer multiorbital models of ${\mathrm{La}}_{4}{\mathrm{Ni}}_{3}{\mathrm{O}}_{10}$}},\ }\href {https://doi.org/10.1103/PhysRevB.110.014503} {\bibfield  {journal} {\bibinfo  {journal} {Phys. Rev. B}\ }\textbf {\bibinfo {volume} {110}},\ \bibinfo {pages} {014503} (\bibinfo {year} {2024}{\natexlab{c}})}\BibitemShut {NoStop}%
\bibitem [{\citenamefont {Kumar}\ \emph {et~al.}(2025)\citenamefont {Kumar}, \citenamefont {Melnick},\ and\ \citenamefont {Kotliar}}]{LNO_cal_JH}%
  \BibitemOpen
  \bibfield  {author} {\bibinfo {author} {\bibfnamefont {U.}~\bibnamefont {Kumar}}, \bibinfo {author} {\bibfnamefont {C.}~\bibnamefont {Melnick}},\ and\ \bibinfo {author} {\bibfnamefont {G.}~\bibnamefont {Kotliar}},\ }\bibfield  {title} {\bibinfo {title} {Softening of $\mathit{dd}$ excitation in the resonant inelastic x-ray scattering spectra as a signature of hund's coupling in nickelates},\ }\href {https://doi.org/10.1103/PhysRevResearch.7.L012066} {\bibfield  {journal} {\bibinfo  {journal} {Phys. Rev. Res.}\ }\textbf {\bibinfo {volume} {7}},\ \bibinfo {pages} {L012066} (\bibinfo {year} {2025})}\BibitemShut {NoStop}%
\bibitem [{\citenamefont {Xie}\ \emph {et~al.}(2024)\citenamefont {Xie}, \citenamefont {Huo}, \citenamefont {Ni}, \citenamefont {Shen}, \citenamefont {Huang}, \citenamefont {Sun}, \citenamefont {Walker}, \citenamefont {Adroja}, \citenamefont {Yu}, \citenamefont {Shen}, \citenamefont {He}, \citenamefont {Cao},\ and\ \citenamefont {Wang}}]{LNO327_neutron_WM}%
  \BibitemOpen
  \bibfield  {author} {\bibinfo {author} {\bibfnamefont {T.}~\bibnamefont {Xie}}, \bibinfo {author} {\bibfnamefont {M.}~\bibnamefont {Huo}}, \bibinfo {author} {\bibfnamefont {X.}~\bibnamefont {Ni}}, \bibinfo {author} {\bibfnamefont {F.}~\bibnamefont {Shen}}, \bibinfo {author} {\bibfnamefont {X.}~\bibnamefont {Huang}}, \bibinfo {author} {\bibfnamefont {H.}~\bibnamefont {Sun}}, \bibinfo {author} {\bibfnamefont {H.~C.}\ \bibnamefont {Walker}}, \bibinfo {author} {\bibfnamefont {D.}~\bibnamefont {Adroja}}, \bibinfo {author} {\bibfnamefont {D.}~\bibnamefont {Yu}}, \bibinfo {author} {\bibfnamefont {B.}~\bibnamefont {Shen}}, \bibinfo {author} {\bibfnamefont {L.}~\bibnamefont {He}}, \bibinfo {author} {\bibfnamefont {K.}~\bibnamefont {Cao}},\ and\ \bibinfo {author} {\bibfnamefont {M.}~\bibnamefont {Wang}},\ }\bibfield  {title} {\bibinfo {title} {{Strong interlayer magnetic exchange coupling in La$_3$Ni$_2$O$_{7+\delta}$ revealed by inelastic neutron scattering}},\ }\href
  {https://doi.org/https://doi.org/10.1016/j.scib.2024.07.030} {\bibfield  {journal} {\bibinfo  {journal} {Science Bulletin}\ }\textbf {\bibinfo {volume} {69}},\ \bibinfo {pages} {3221} (\bibinfo {year} {2024})}\BibitemShut {NoStop}%
\end{thebibliography}
\end{document}